\documentclass[letterpaper,twocolumn,10pt]{article}
\usepackage{usenix-2020-09}

\usepackage{tikz}
\usepackage{amsmath}
\usepackage{amssymb}
\usepackage{filecontents}
\usepackage{algorithm}
\usepackage{algorithmic}
\usepackage{ulem}
\usepackage{enumitem}
\usepackage{xspace}
\usepackage{booktabs}     
\usepackage{multirow}    
\usepackage{xcolor}    
\usepackage{graphicx}   
\usepackage{subcaption}
\usepackage{tabularx}

\usepackage{subcaption}

\usepackage{tikz}
\usepackage{pgfplots}
\usepgfplotslibrary{polar}
\usepackage{pifont}
\pgfplotsset{compat=1.18}

\setlist[itemize]{leftmargin=1em, itemsep=0pt, parsep=0pt, topsep=2pt}

\newcommand\blfootnote[1]{%
  \begingroup
  \renewcommand\thefootnote{}\footnote{#1}%
  \addtocounter{footnote}{-1}%
  \endgroup
}
\newcommand{\sys}{\textsf{Morphe}\xspace}

\begin{document}

\date{}

\title{\Large \bf \sys: High-Fidelity Generative Video Streaming with Vision Foundation Model}


\author{
Tianyi Gong$^{1,2}$, Zijian Cao$^{1,2}$, Zixing Zhang$^{1,2}$, Jiangkai Wu$^{3}$, Xinggong Zhang$^{3}$, \\Shuguang Cui$^{1,2}$, Fangxin Wang$^{1,2}$\\
$^{1}$\textit{School of Science and Engineering, The Chinese University of Hong Kong, Shenzhen}, \\\quad $^{2}$\textit{Shenzhen Future Network of Intelligence Institute},
\\\quad $^{3}$\textit{Wangxuan Institute of Computer Technology, Peking University}
}
 \maketitle
 \blfootnote{The first three authors, Tianyi Gong, Zijian Cao, and Zixing Zhang, contributed equally to this paper. Fangxin Wang is the corresponding author.}

\vspace{-2.5em}
\begin{abstract}

Video streaming is a fundamental Internet service, while the quality still cannot be guaranteed especially in poor network conditions such as bandwidth-constrained and remote areas. Existing works mainly work towards two directions: traditional pixel-codec streaming nearly approaches its limit and is hard to step further in compression; the emerging neural-enhanced or generative streaming usually fall short in latency and visual fidelity, hindering their practical deployment. 

Inspired by the recent success of vision foundation model (VFM), we strive to \textit{harness the powerful video understanding and processing capacities of VFM to achieve generalization, high fidelity and loss resilience for real-time video streaming with even higher compression rate.} We present , the first revolutionized paradigm that enables VFM-based end-to-end generative video streaming towards this goal. Specifically, \sys\footnote{Our project page can be available at \url{https://inml-tygong.github.io/Morphe}} employs joint training of visual tokenizers and variable-resolution spatiotemporal optimization under simulated network constraints. Additionally, a robust streaming system is constructed that leverages intelligent packet dropping to resist real-world network perturbations. Extensive evaluation demonstrates that \sys achieves comparable visual quality while saving 62.5\% bandwidth compared to H.265, and accomplishes real-time, loss-resilient video delivery in challenging network environments, representing a milestone in VFM-enabled multimedia streaming solutions.


\end{abstract}

\section{Introduction}
\vspace{-0.5em}

Nowadays Video content dominates the Internet traffic, accounting for more than 65\% based on \cite{sandvine2023global}. However, accessing high-quality video content is often hindered by insufficient and unstable network bandwidth between client and server, which is even worse for high-resolution live video streaming. 
In such network-constrained scenarios as high-speed railway and remote areas, people's demand for premium media content remains difficult to satisfy.


\noindent\textbf{The Main Roadmap:} 
To tackle this problem, multimedia streaming employs codec optimization and adaptive bitrate control, and network-aware protocols to maintain video quality under constrained conditions.
\textit{Traditional methods} (e.g., codec with H.264/265 \cite{wiegand2003overview,sze2014high} with adaptive bitrate\cite{spiteri2020bola}) focus on pixel-level encoding and compression, pursuing pixel-perfect reconstruction. However, when confronted with insufficient or fluctuating bandwidth, they struggle to maintain optimal performance, leading to compression artifacts or playback interruptions. While widely adopted in modern streaming service, pixel-based codec methods nearly approach their physical limits in bandwidth-constrained environments where visual quality compromises become necessary.


\textit{Neural-enhanced streaming}\cite{kim2020neural,yeo2020nemo,jeong2024real} emerges as a promising direction, which leverages learning-based post-processing to achieve substantial bandwidth savings, but often suffers from considerable computational overhead as well as increased latency. \textit{Generative streaming approaches}\cite{wang2024maskgct,artioli2024generative} attempt to reconstruct video content from compact semantic representations, essentially generating frames from high-level descriptions rather than transmitting pixels\cite{zhou2024survey}. These methods offer promising bandwidth efficiency but often suffer from unacceptable visual quality degradation, typically stemming from poor generalizability. Moreover, due to the dense mapping between semantics and pixels and the lack of recovery mechanism, a slight packet loss can lead to serious distortion. 


The core tenet of video streaming lies in \textbf{streaming high-fidelity, real-time, and loss-resilient video content with highest possible video compression}. However, under complex network conditions, existing streaming systems—both traditional pixel-based codecs and emerging neural approaches—struggle to balance this triangular relationship effectively. Compromises in any dimension continue to severely degrade Quality of Service (QoS) across all current streaming paradigms.


\noindent\textbf{Opportunities and Challenges:} Vision Foundation Models (VFMs)\cite{bommasani2021opportunities} have attracted considerable attention owing to their exceptional capabilities in video understanding\cite{yuan2021florence} and generation\cite{wu2024vila}. 
VFMs present a transformative opportunity for content-aware video streaming systems through their unique "understanding before generation" paradigm. Pre-trained on large-scale visual corpora via unsupervised learning, VFMs excel at distilling redundant visual data into compact semantic representations that naturally align with human perceptual priorities—prioritizing faces, salient objects, and textual elements over low-level textural details. This semantic-aware encoding enables extraordinary compression gains (exceeding $300\times$ in recent studies~\cite{phung2025exploring}) while maintaining perceptually-optimized quality beyond conventional pixel-level metrics. Moreover, the zero-shot generalization capability inherent to VFMs ensures robust performance when confronted with real-world data variability. Inspired by the remarkable potential of VFMs, we are motivated to investigate a question: \textit{Can we achieve Vision Foundation Model-based video coding towards high-fidelity, loss-resilient, and real-time streaming service?}


Unfortunately, through our analysis in \S\ref{nextgvs}, applying VFM for video streaming still faces triple challenges:

\noindent \textbf{C1: How to maximize VFM-based compression with high-fidelity visual recovery?}
VFMs are fundamentally designed for semantic understanding and generation tasks, primarily focusing on low-frequency content, which inherently conflicts with high-fidelity video delivery that demands detailed texture preservation . Beyond severe detail loss, spatiotemporal discontinuities also pose significant challenges when adapting VFMs to video streaming, potentially stemming from high spatiotemporal compression ratios and improper video chunking strategies. The fundamental mismatch between semantic intelligence and high-fidelity transmission requirements explains why simply migrating VFMs to streaming scenarios fails to reconstruct photorealistic video content.

\noindent \textbf{C2: How to guarantee low-latency streaming with computationally intensive VFMs?}
VFMs typically demand substantial computational overhead for tokenization and inference, creating inherent tension with real-time streaming requirements. The computational bottleneck primarily stems from complex encoder-decoder architectures that process visual inputs through multi-stage operations including 3D Haar wavelet transforms, causal residual blocks, and causal spatio-temporal attention mechanisms\cite{agarwal2025cosmos}. Directly applying VFMs to high-resolution video streams exacerbates this issue, as the sequential processing nature and attention complexity create significant processing latency.

\noindent \textbf{C3: How to ensure loss-resilient and network-adaptive streaming with VFM-based codec?}
Traditional streaming systems rely on redundancy-based error correction, but VFM-based systems face unique vulnerabilities due to their semantic compression paradigm. When network conditions fluctuate dramatically, excessive bandwidth leads to underutilization, while insufficient bandwidth may cause packet losses. Packet losses in semantically compressed streams are particularly devastating, as they can trigger catastrophic quality degradation and temporal inconsistencies that propagate across frames. Current VFMs lack error recovery mechanisms and network-aware adaptation capabilities that can dynamically adjust compression strategies based on channel conditions.

\begin{table}[t]
  \footnotesize \centering
  \caption{\label{tab:streaming_paradigms} Comparison of Different Streaming Paradigms.}
  \renewcommand{\arraystretch}{0.9}
  \begin{tabular}{lccc}
    \toprule
    \textbf{Technical Paradigm} & \textbf{Fidelity} & \textbf{Efficiency} & \textbf{Robustness} \\
    \midrule
    Traditional Codec \cite{wiegand2003overview} \cite{sze2014high} & Low & High & High \\
    Neural Enhanced \cite{yeo2018neural} & Medium & Medium & High \\
    Generative Neural Codecs \cite{cheng2024grace}& Medium & Medium & High \\
    Diffusion-based  \cite{wu2024promptus}& Medium & Low & Low \\
    VFM-based \sys & High & High & High\\
    \bottomrule
  \end{tabular}
\end{table}

\noindent\textbf{Design and Contribution:} To address the aforementioned challenges, we propose \sys\footnote{\textsf{Morphe} derives from the Greek deity of dreams who crafted and shaped human forms. This indicates the core function: to intelligently shape and reconstruct high-fidelity video from highly compressed latent features.}, an innovative end-to-end video streaming system empowered by VFMs. To the best of our knowledge, \sys stands out as the first system to leverage vision foundation models to achieve high-fidelity, low-latency, and loss-resilient advanced video service, as highlighted again in the Table \ref{tab:streaming_paradigms}. The core of our design lies in three aspects: a generative codec fine-tuned from VFM that simultaneously ensures high compression ratios and high fidelity, a resolution scaling accelerator with neural-enhanced processing that significantly reduces computational complexity, and an adaptive streaming controller that responds to network fluctuations. Specifically, \sys incorporates the following three innovative design components to efficiently adapt VFMs for multimedia networking.
\begin{itemize}

\item \textbf{Visual-enhanced Generative Codec}: To address \textit{C1}, we design an asymmetric spatiotemporal compression strategy that fine-tunes Vision Foundation Models (VFMs) and minimizes visual artifacts during encoding-decoding. This module automatically optimizes compression efficiency while preserving visual quality through advanced generative reconstruction and inter-frame smoothing, achieving unprecedented video fidelity at extreme compression ratios with photorealistic playback quality.

\item \textbf{Resolution Scaling Accelerator}: To address \textit{C2}, we propose a comprehensive acceleration framework that can significantly improve system efficiency. This module automatically balances computational complexity and visual quality through intelligent downsampling during preprocessing and deep learning-based super resolution during post-processing, enabling real-time 1080p streaming under resource constraints.

\item \textbf{Network-Adaptive Streaming Controller}: To address \textit{C3}, we establish a robust streaming architecture with scalable bitrate control and recovery mechanisms towards network dynamics and bandwidth limitations. This module automatically adapts streaming parameters to network conditions through adaptive bitrate algorithms and self error correction strategies, ensuring seamless video delivery even under severe packet loss conditions. 
\end{itemize}

\sys reduces bitrate by 62.5\% compared to H.265 while maintaining comparable visual quality. The system achieves 65 fps real-time streaming on a server with a single NVIDIA RTX 3090 and 94.2\% bandwidth utilization in real network transmission.The contributions of this paper are summarized as follows:
\begin{itemize}
\item We identify the key challenges of applying VFMs to video streaming systems and demonstrate that existing compression and streaming approaches are inadequate for achieving high-fidelity, real-time,loss-resilient video delivery under network constraints (\S\ref{bacground}).
\item We then design \sys, the first VFM-powered end-to-end video streaming system that incorporates temporalspatial visual enhancement modules for improved coedc quality, comprehensive preprocessing and postprocessing optimizations for system efficiency, and an intelligent streaming controller to handle network fluctuations (\S\ref{model design}, \S\ref{downsample}, \S\ref{network}).
\item Comprehensive experiments and simulations are provided to demonstrate the system's applicability and scalability, showing superior quality, real-time HD streaming under bandwidth constraints, and robust performance despite packet loss under challenging environments (\S\ref{evaluation}).
\end{itemize}

\vspace{-1em}
\section{Background and Motivation}\label{bacground}\vspace{-0.5em}

\subsection{Bandwidth-starved Multimedia Streaming}

\textbf{The Bandwidth Concern in Global Video Communication.}
Consider business travelers in rural areas joining critical video conferences through degraded 2G networks supporting only 100 kbps—far below the 300 kbps minimum for intelligible video calls. Despite video communication's ubiquity, billions remain excluded from high-quality visual connectivity due to severe bandwidth limitations. 

\begin{figure}[t]
    \centering
    \begin{subfigure}{0.23\textwidth}
        \centering
        \includegraphics[width=\textwidth]{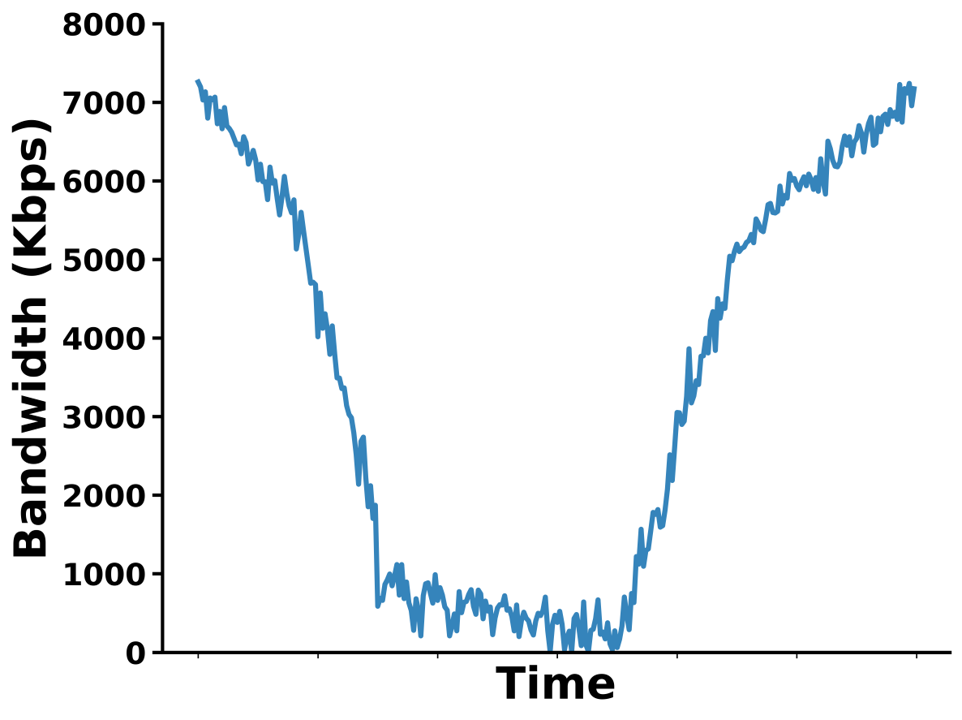}
        \caption{Network trace during train travel (through tunnels)}
        \label{fig:left}
    \end{subfigure}
    \hfill  
    \begin{subfigure}{0.23\textwidth}
        \centering
        \includegraphics[width=\textwidth]{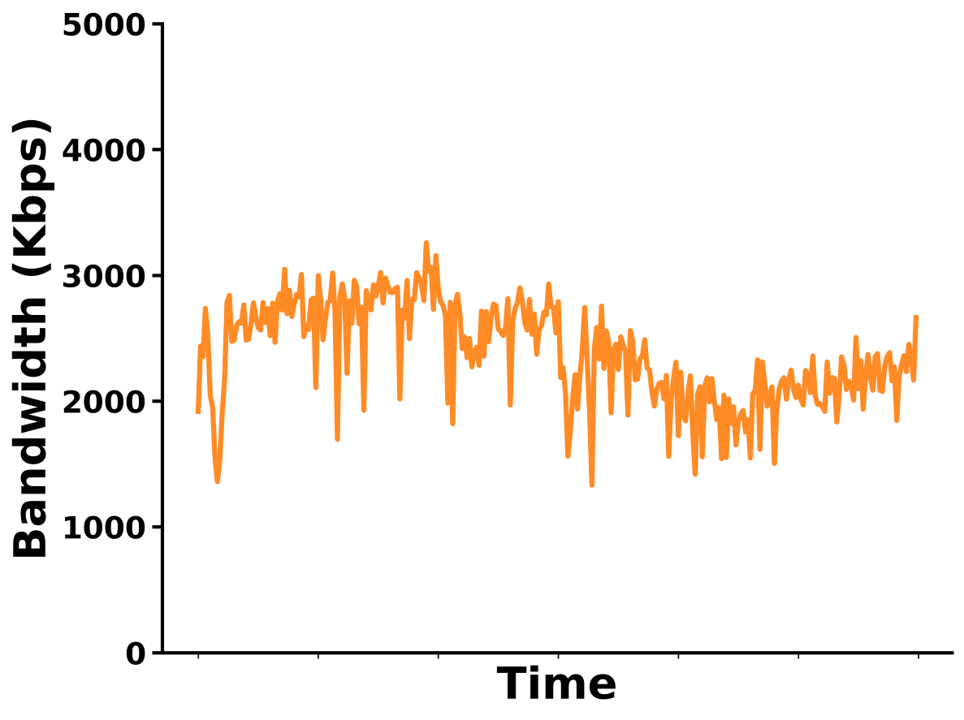}
        \caption{Network trace during countryside self-driving tours}
        \label{fig:right}
    \end{subfigure}
    \vspace{-2mm}
    \caption{Case Study of Bandwidth-Constrained Multimedia Communication.}
    \label{fig:intro}
\end{figure}

This constraint extends beyond developing regions. Emergency responders require crystal-clear video feeds through compromised disaster networks. Remote industrial facilities demand real-time equipment monitoring despite limited connectivity. High-speed transportation systems face intermittent connections, yet passenger safety depends on uncompromised video quality. As illustrated in the Figure \ref{fig:intro}, our investigation and collection of network traces from high-speed rail journeys and vehicle driving in edge areas reveal that many real-world scenarios still suffer from harsh network conditions. These scenarios share a common challenge: delivering high quality video content under extreme bandwidth constraints.

\vspace{-1em}

\subsection{Traditional Video Streaming}

\textbf{Exhausted Performance Under Low Bandwidth.} Current video coding technologies reveal architectural limitations when pushed to extreme bandwidth constraints. Traditional codecs like H.264\cite{wiegand2003overview} and H.265\cite{sze2014high} employ pixel-based intra-frame coding and inter-frame prediction paradigms optimized for moderate compression ratios. These codecs achieve compression primarily by eliminating spatial redundancy through correlation between adjacent pixel blocks and temporal redundancy through inter-frame prediction techniques that exploit similarity between consecutive frames. However, this redundancy-removal approach encounters a fundamental ceiling—the information-theoretic physical constraints that bound achievable compression ratios.

Neural-embedded codecs\cite{djelouah2019neural,li2024neural,lin2020m,lu2019dvc} improve upon traditional methods by using neural networks to learn complex motion patterns and correlations, but they remain constrained by the conventional redundancy elimination framework. This architectural limitation bounds their potential, as they still rely on pixel-level reconstruction assumptions. Below 50 kbps, these approaches fail catastrophically, producing videos with severe blocking artifacts, ghosting, and detail loss that make them unusable for practical viewing.

\vspace{-1em} 
\subsection{Learning-based Video Streaming}

\subsubsection{Neural-Enhanced Streaming}
\textbf{Insufficient Learning and Limited Generalizability.} Neural-enhanced video compression leverages deep learning to achieve superior rate-distortion trade-offs by transmitting highly distorted low-bitrate videos and restoring them using neural network post-processing. Unlike traditional redundancy-based approaches, this method actively discards most information, relying on neural networks' powerful restoration capabilities to recover lost details while achieving substantial compression ratios.

However, these AI-driven networks face significant vulnerabilities due to domain gaps between training datasets and real-world deployment scenarios\cite{yang2021real}. Performance degrades with out-of-distribution content, introducing artifacts like unnecessary high-frequency details and hallucination effects that compromise authenticity. While solutions like NAS\cite{kim2020neural,yeo2020nemo,yeo2018neural} propose fine-tuning networks for each video segment, transmitting these adapted models increases bitrate, creating a fundamental trade-off between adaptation capability and bandwidth efficiency.

\begin{figure}[t]
    \centering
    \includegraphics[width=0.48\textwidth]{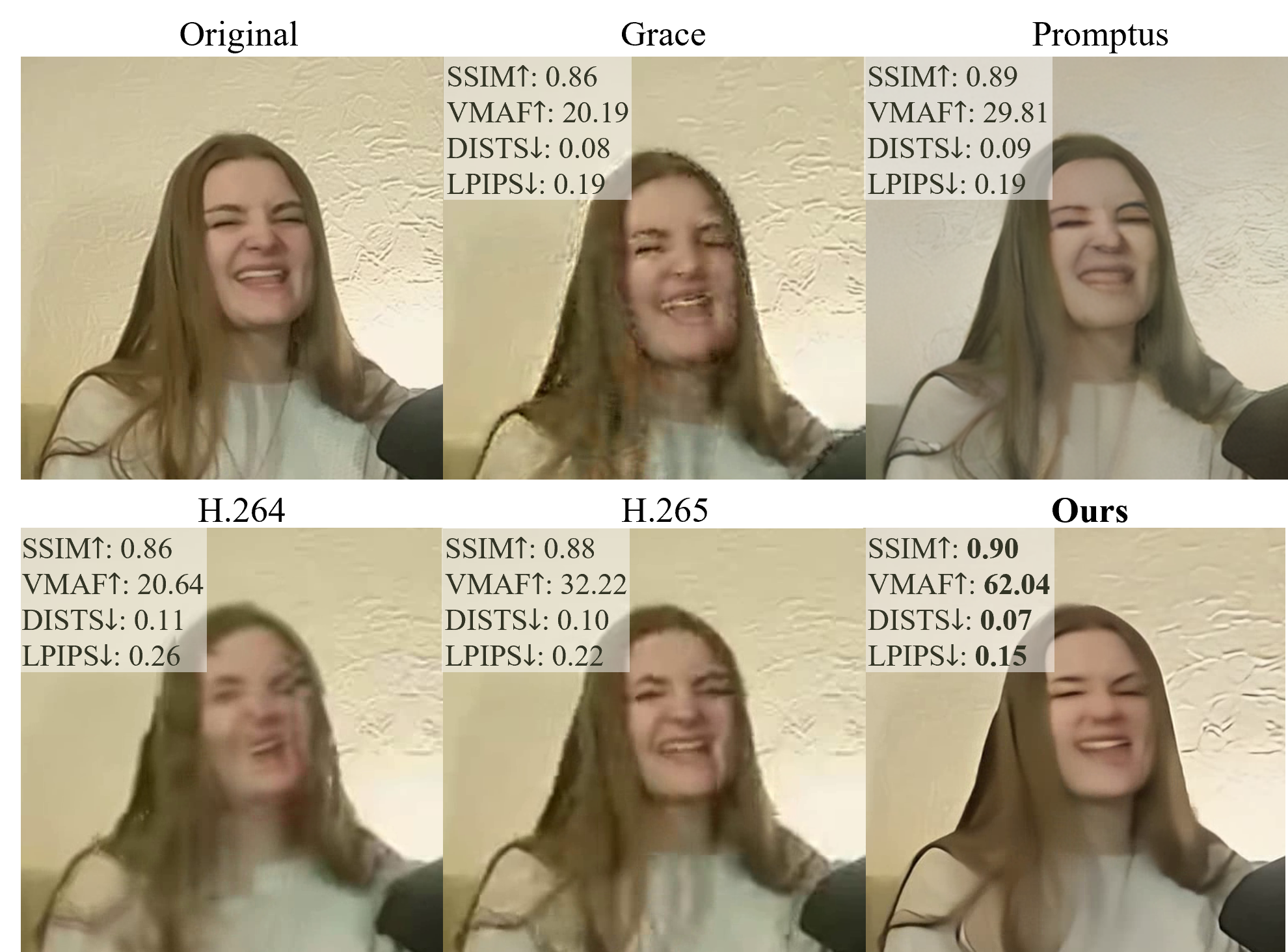}
    \caption{Visual perception of different streaming technologies at 400 kbps . Most existing solutions show severe artifacts under poor network conditions.}
    \label{fig:intro compare}
    \vspace{-1.7em}
\end{figure}

\vspace{-1em}
\subsubsection{Loss-Resilient Generative Codecs}

The evolution of loss-resilient generative codecs has progressed through distinct technological paradigms. Early systems like GRACE\cite{cheng2024grace} pioneered dropout-based training but relied on oversimplified models assuming uniform random packet loss, degrading under real network conditions with temporal clustering.Token-based approaches like Reparo\cite{li2023reparo} advanced the field using VQ-GAN\cite{esser2021taming} codebooks to map video content into discrete indices, achieving dramatic bitrate reductions while maintaining perceptual quality. Beyond the aforementioned weak generalization and task-specific limitations, this approach introduces other fundamental problems:


\noindent\textbf{Malfunctioning Motion Modeling.} Systems like GRACE model frames independently, resulting in weak temporal consistency that generates severe mosaic artifacts around motion regions under extreme compression. This frame-independent modeling approach, rather than considering the video holistically, leads to catastrophic error propagation - a classic video transmission problem where corrupted frames cascade through temporal dependencies. 

\noindent\textbf{Frame Resolution Inflexibility.} Token-based systems exhibit strict resolution dependencies tied to training configurations. Models trained on 720p content cannot accommodate higher-resolution inputs like 1080p without complete retraining of codebooks and networks. This inflexibility limits deployment scalability, necessitating multiple specialized models for different resolutions and increasing system complexity.

\vspace{-1em}
\subsubsection{Diffusion-based Generative Streaming}

Diffusion-based approaches for video streaming have demonstrated significant potential, with DiffusionStream\cite{kodaira2023streamdiffusion} establishing feasibility for real-time transmission. Systems like Promptus\cite{wu2024promptus} advance this paradigm using compact semantic prompts to guide diffusion-based generation, achieving remarkable generalization across diverse content types and arbitrary resolutions. However, diffusion-based streaming faces two critical limitations:

\noindent\textbf{Imperfect Efficiency-Quality Trade-offs.} Diffusion models rely on extensive denoising steps, requiring sophisticated optimization and complex hardware acceleration to approach real-time performance. Achieving superior visual quality demands prolonged training periods, restricting deployment to offline playback scenarios. Additionally, diffusion controllability remains weak, frequently producing AI artifacts—temporal inconsistencies or unrealistic textures—that are easily detectable and fall short of photorealistic standards required for professional applications.

\noindent\textbf{Poor Network Resilience.} Existing diffusion-based designs ignore fluctuating network conditions, lacking robust packet loss resilience. The sequential nature of diffusion generation creates vulnerabilities where prompt corruption or incomplete transmission cascades into complete frame reconstruction failures. Diffusion Model-based systems struggle to maintain semantic coherence when prompt information is partially lost, making them unsuitable for unreliable network environments typical of mobile streaming.

Along the development of video streaming, we continue to witness many limitations. As shown in Figure \ref{fig:intro compare}, current streaming systems suffer significant video quality degradation when facing insufficient bandwidth and network jitter.

\vspace{-1em}
\subsection{Advancing Vision Foundation Models Towards High-Quality Video Streaming}\label{nextgvs}

\begin{figure}[t]
    \centering
    \includegraphics[width=0.5\textwidth]{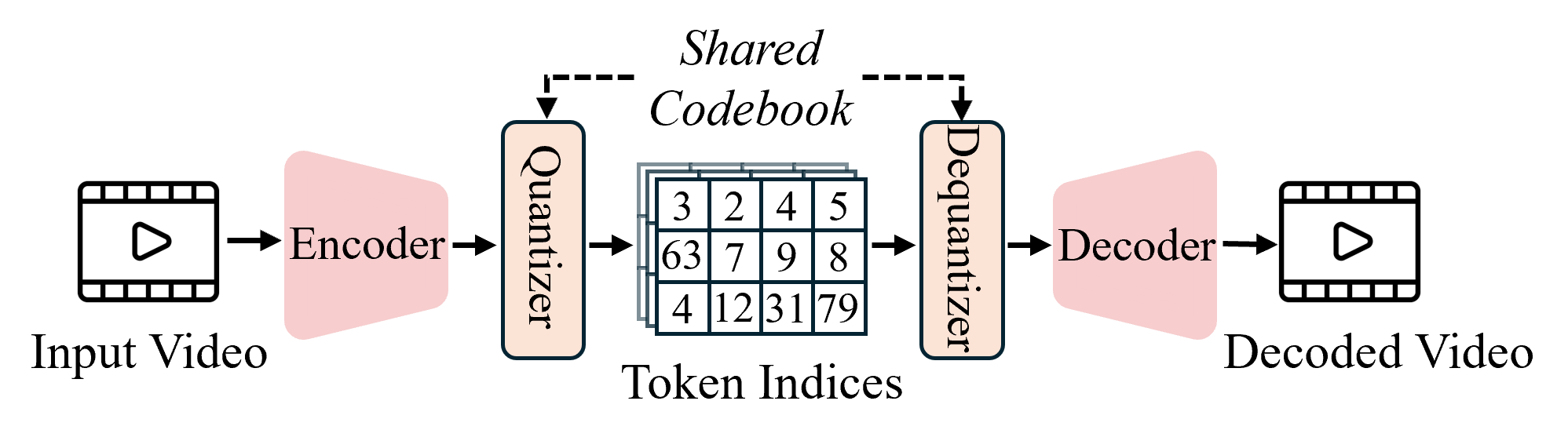}
    \vspace{-5mm}
    \caption{Video tokenization pipeline in Vision Foundation Models. An input video is encoded into tokens, which are usually much more compact than the input video. The decoder then reconstructs the input video from the tokens. }
    \label{fig:ori token}
    \vspace{-1em}
\end{figure}


Vision Foundation Models (VFMs) have emerged as transformative architectures offering unprecedented semantic understanding and generalizability for video compression. These models employ encoder-decoder architectures that compress visual data while preserving semantic content, as illustrated in Fig.\ref{fig:ori token}. The encoding function $f_{\phi}$ transforms input videos $x_{0:T} \in \mathbb{R}^{(1+T) \times H \times W \times 3}$ into compact latent representations $z_{0:T'} \in \mathbb{R}^{(1+T') \times H' \times W' \times C}$ through multi-dimensional downsampling with spatial compression factors $s_{HW}$ and temporal compression factor $s_T$, which are reconstructed via decoder $g_{\theta}$. This architecture enables semantic compression with significant gains based on perceptual importance.


However, directly transplanting VFMs to multimedia streaming is not feasible. The fundamental tension between VFMs' understanding-oriented design and streaming's quality-centric demands reveals deep-rooted incompatibilities. These gaps, as outlined below, represent the core challenges we must address:

\begin{table}[t]
\centering
\vspace{-1mm}
\caption{Comparative Analysis of Vision Foundation Models for Video Encoding and Decoding.}
\renewcommand{\arraystretch}{0.9}
\setlength{\tabcolsep}{6pt}
\small
\begin{tabular}{lccc}
\toprule
\textbf{Model} & \textbf{Precision} & \textbf{Enc.(FPS)} & \textbf{Dec.(FPS)} \\
\midrule
VideoVAE Plus \cite{xing2024large} & fp16 & 2.12 & 1.47 \\
Cosmos \cite{agarwal2025cosmos} & fp16 & 6.21 & 5.08 \\
CogVideoX-VAE \cite{yang2024cogvideox} & fp16 & 5.52 & 1.95 \\
\bottomrule
\end{tabular}
\label{tab:video_vae_performance}
\vspace{-1.5em}
\end{table}

\noindent\textbf{High Fidelity versus Perceptually Lossy VFMs.}
Video streaming demands photorealistic playback experiences, yet VFMs primarily focus on video understanding and generation, where low-frequency content representation plays a crucial role, inevitably losing substantial high-frequency details essential for visual fidelity. Current VFM tokenizers fundamentally sacrifice high-quality detail for semantic understanding, creating an unbridgeable gap for high-fidelity streaming applications. Furthermore, the lossy compression inherent in VFM tokenizers leads to perceptual artifacts—spatial blurring, temporal flickering, and phantom details that violate human visual perception expectations. Persistent visual distortions and poor frame-to-frame consistency result in uncomfortable viewing experiences.

\noindent\textbf{Low Latency versus Computationally Intensive VFMs.}
Video streaming requires seamless and smooth playback experiences, yet VFMs are inherently computationally intensive designs that fundamentally contradict real-time streaming requirements. Our investigation reveals that VFM processing times for high-resolution video content are prohibitively high for real-time applications. The inference latency of these AI-native architectures creates insurmountable bottlenecks when processing 1080p streams at standard frame rates, as demonstrated by our investigation shown in Table \ref{tab:video_vae_performance}. This establishes an inherent incompatibility between VFM computational demands and the real-time constraints essential for seamless video streaming.

\noindent\textbf{Loss Resilience versus Network-Agnostic VFMs.}
Video streaming systems must adapt to challenging network environments with severe packet loss, demanding sophisticated quality control to maintain QoS. However, current VFMs completely lack network-awareness capabilities, operating as isolated black boxes without network condition integration. Packet loss during transmission can catastrophically corrupt semantic tokens, leading to complete reconstruction failures rather than graceful quality degradation. To enable VFM-based systems to survive real-world networks, loss-resilient strategies and network-adaptive tokenization mechanisms are essential, yet VFMs fundamentally lack these prerequisites for robust streaming systems.

\vspace{-3mm}

\section{\sys Overview}\label{overview}
\vspace{-0.5em}


In response to the above challenges and insights, we design three modules for \sys: Visual-enhanced Generative Codec (\S\ref{model design}), Resolution Scaling Accelerator (\S\ref{downsample}), and Network-Adaptive Streaming Controller (\S\ref{network}). The workflow of \sys is illustrated in the Figure \ref{fig:system}, with specific details listed as follows:

\noindent\textbf{Visual-enhanced Generative Codec (VGC)} performs high-fidelity compression on preprocessed videos through residual codec, temporal smoothing, and intelligent token dropping. Unlike traditional VFM models, VGC fine-tuned to focus on visual quality keeps important visual details and allows flexible control of bitrate. The codec generates variable-rate data streams that can be dynamically adjusted based on NASC's network feedback, achieving extreme compression ratios with minimal degradation.

\noindent\textbf{Resolution Scaling Accelerator (RSA)} dynamically selects the appropriate resolution for VGC encoding based on bandwidth to balance quality and latency, enabling real-time 1080p streaming under resource constraints. The preprocessed downsampled frames are sent to VGC for compression, for decoded streams, a neural super-resolution network jointly leverages VGC's decoder features to restore full resolution.

\noindent\textbf{Network-Adaptive Streaming Controller (NASC)} orchestrates the streaming of VGC-compressed tokens across fluctuating network conditions. NASC achieves scalable bitrate control by jointly managing VGC's bitrate adjustment and provides robust streaming protocols with self-recovery mechanisms. When bandwidth varies or packet losses occur, NASC prioritizes critical semantic tokens and reconstructs missing information through temporal-spatial correlations, ensuring stable video delivery even under lossy network conditions.

\vspace{-3mm}

\section{Visual-enhanced Generative Codec}\label{model design}
\vspace{-0.5em}

\begin{figure*}[t]
  \centering
  \includegraphics[width=0.9\textwidth]{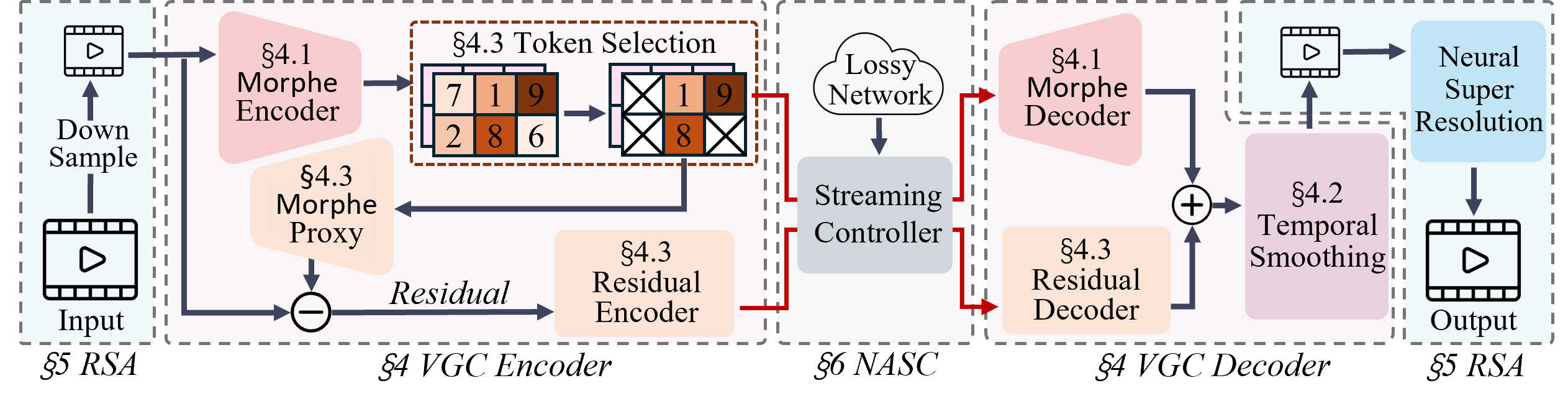}
  \vspace{-2mm}
  \caption{\sys Overview. It consists of three core modules: Visual-enhanced Generative Codec (\S\ref{model design}), Resolution Scaling Accelerator (\S\ref{downsample}), and Network-Adaptive Streaming Controller (\S\ref{network}).} 
  \label{fig:system}
  \vspace{-1em}

\end{figure*}

As shown in Figure \ref{fig:system} VGC encoder and decoder, this section explains VGC codec in three parts. We begin with the overall architecture of the codec model (\S\ref{codec}), then describe how we suppress video jitter and improve stability (\S\ref{overlap}), and finally present how to implement scalable video encoding to achieve adjustable compression ratios (\S\ref{drop and res}).

\vspace{-1em}
\subsection{Generative Codec Architecture}\label{codec}

To balance high compression ratio with visual quality, VGC employs a multi-interface codec framework with asymmetric compression. This section first presents the codec architecture and then describes the multi-interface framework.

\noindent\textbf{Video Generative Codec of VGC.} 
Original video foundation models are primarily designed to extract semantic features of video for subsequent video generation or understanding. To serve diverse downstream tasks, they commonly offer two standard compression settings: (1) $8\times$ temporal downsampling with $16\times16$ spatial downsampling, which yields a high compression ratio but introduces noticeable temporal jitter and loss of spatial detail. And (2) $4\times$ temporal downsampling with $8\times8$ spatial downsampling, which preserves more detail and temporal smoothness but achieves a lower compression ratio that is ill-suited to bandwidth-constrained networks. This fundamental problem between compression and visual quality raises a key design question: \textit{Can we find a balance that preserves critical spatial detail, minimizes temporal jitter, and still fits within narrowband transmission?}

From a systematic analysis of streaming video traffic, we find that the data required to increase spatial detail far exceed that needed to maintain temporal smoothness. This insight guides our design: prioritize spatial detail and allocate more compression to the temporal dimension. Concretely, we propose an asymmetric configuration that keeps $8\times8$ spatial compression while increasing temporal compression to $8\times$. This choice maintains a high compression ratio while preserving high-frequency spatial details to meet high-fidelity video demands over narrow links. Although stronger temporal compression will introduce jitter, the temporal smoothing enhancement techniques in \S\ref{overlap} effectively mitigate this issue, enabling \sys to optimize the trade-off between compression efficiency and perceptual quality.

\noindent\textbf{Multi-objective Optimization Framework.} To enhance video reconstruction quality and system efficiency, we incorporate a multi-interface framework into VGC featuring two critical interfaces: an internal gradient interface for iterative optimization and an external data interface for visual enhancement. The gradient interface enables adaptive training objectives across different stages, while the external interface integrates visual enhancement modules and network control components. VGC's architecture supports joint scheduling for visual quality optimization and network adaptation, providing unified control for optimization algorithms and ensuring high-quality streaming under challenging network conditions.

\vspace{-1em}
\subsection{Temporal Consistency Enhancement}\label{overlap}


To mitigate jitter caused by aggressive temporal compression, VGC designs a pixel blending scheme that leverages codec co-optimization. It combines model optimization with inter-frame smoothing to suppress temporal artifacts.

\noindent\textbf{Pixel Blending and Codec Optimization.} Inspired by conventional video codecs, VGC adopts Group-of-Pictures (GoP) for segment-based encoding and decoding. However, the combination of per-GoP processing and strong temporal compression inevitably causes temporal jitter at GoP boundaries. A straightforward idea is to blend pixels across adjacent GoPs near their boundary to achieve smooth transitions. However, since VGC encodes each GoP independently, spatial misalignment and brightness inconsistencies arise. Direct blending under these conditions often worsens artifacts and blurring instead of improving visual coherence.

To resolve this, VGC uses the internal gradient interface introduced in \S\ref{codec} and proposes an iterative codec co-optimization scheme for boundary-frame blending. The scheme proceeds in two stages.

First, to exploit priors in existing vision foundation model and suppress flicker, we add a dedicated constraint tailored to GoP boundary alignment. Specifically, we compute a pixel-level loss over the boundary region:
\begin{equation}
L_{\text{temporal}} \;=\; \frac{1}{n}\sum_{i=1}^{n} \left\lVert \hat{x}_{\text{curr},i} - x_{\text{prev},T-n+i} \right\rVert
\end{equation}
where $\hat{x}_{\text{curr},i}$ denotes the first $n$ reconstructed frames of the current GoP and $x_{\text{prev},T-n+i}$ denotes the last $n$ frames of the previous GoP. While this pixel-difference supervision does not fully capture temporal coherence, it effectively exploits model priors and reduces training cost. With this constraint, frames near the GoP boundary become closer in pixel space.

Then blending is applied to the optimized boundary region:
\begin{equation}
\hat{x}_{\text{blend},i} \;=\; \alpha_i \cdot \hat{x}_{\text{prev},T-n+i} \;+\; (1-\alpha_i)\cdot \hat{x}_{\text{curr},i}
\end{equation}
where $\alpha_i$ is the linear interpolation weight, which we set as $\alpha_i = \frac{n-i}{n}$. Because training enforces pixel proximity at GoP boundary, this smoothing yields a seamless transition from the previous GoP to the current one. Using this smoothing mechanism, VGC significantly improves temporal smoothness without incurring any additional transmission cost. 

\subsection{Scalable Video Coding}\label{drop and res}



\begin{figure}[t]
    \centering
    \includegraphics[width=0.45\textwidth]{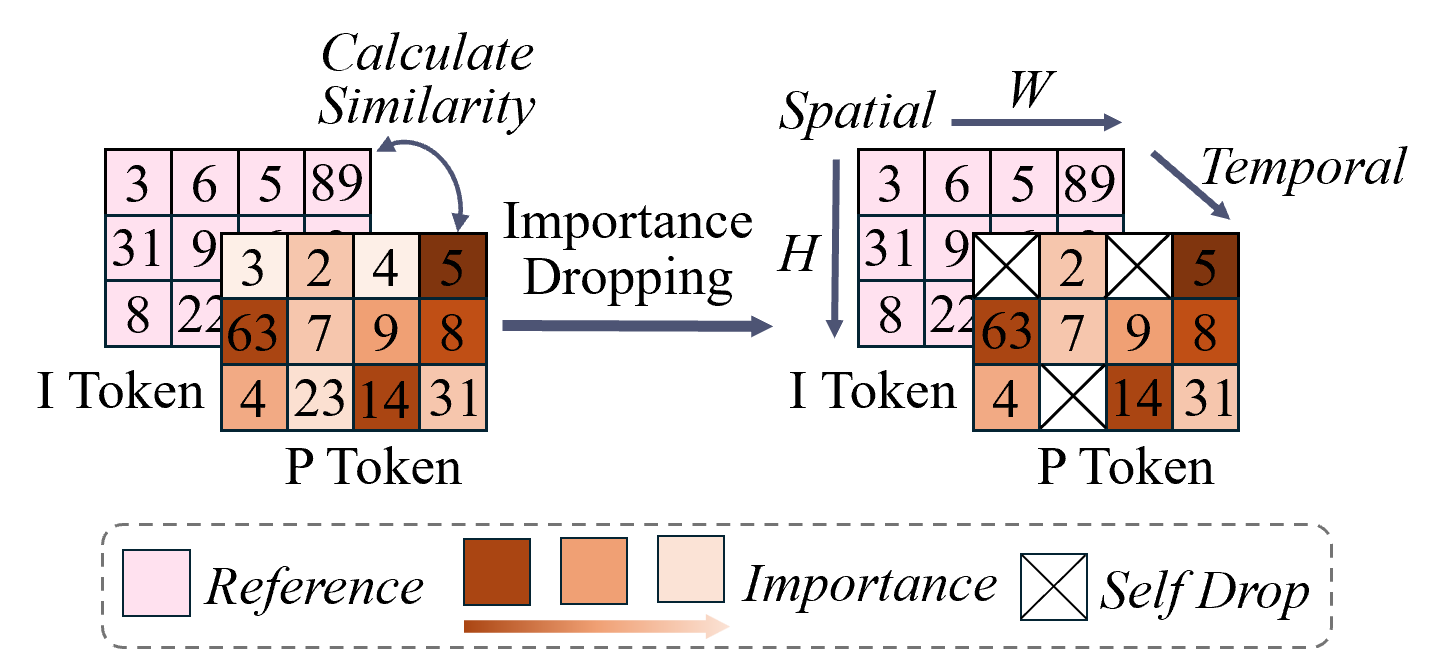}
    \caption{ Illustration of Similarity-based Token Selection.}
    \label{fig:selection}
    \vspace{-2.5mm}
\end{figure}

To expand the operating bitrate range and visual quality of codec, \sys VGC employs scalable video encoding. In this section, we first introduces similarity-based token selection, then describes pixel residual computation and transmission.

\noindent\textbf{Similarity-based Token Selection.} In low-bandwidth network, even aggressive codec compression may exceed available bandwidth. A straightforward way is to randomly drop semantic tokens, but this ignores semantic importance, risking critical visual content loss and quality degradation. Inspired by I/P frame structures in traditional codecs, we observe significant redundancy within GoP semantic representations. This enables further compression: by identifying and selectively discarding similar tokens, we can preserve the transmission of core visual content even under very limited bandwidth.

VGC adopts an I/P-style structure like traditional codecs. We set each GoP contains 9 frames. The first frame serves as a reference I frame and is compressed only spatially, yielding a full semantic token matrix. The subsequent 8 frames (P frames) are jointly compressed in spatial and temporal at $8\times$, producing another token matrix. This design naturally provides a reference baseline for token importance estimation.

As shown in Figure \ref{fig:selection}, for the P-frame token matrix, we evaluate the importance of each token by its similarity to the I-frame tokens at the same spatial location. Concretely, for the $k$-th GoP, let $T_k^{P}$ denote the token matrix from the 8 jointly compressed frames and $T_k^{I}$ the token matrix from the first (I) frame. We compute a per-token cosine similarity:
\begin{equation}
S_{i,j} \;=\; \frac{T_k^{P(i,j)} \cdot T_k^{I(i,j)}}{\lVert T_k^{P(i,j)} \rVert \, \lVert T_k^{I(i,j)} \rVert}
\end{equation}
where $T_k^{P(i,j)}$ and $T_k^{I(i,j)}$ are the token vectors at location $(i,j)$ in the P and I token. A higher $S_{i,j}$ indicates greater temporal redundancy relative to the reference; such tokens can be deprioritized and, under bandwidth pressure, dropped first.

Given the computed similarities, the system selects a dynamic threshold $\tau$ based on the current available bandwidth and marks tokens with $S_{i,j} > \tau$ as discardable. This cosine-based scoring is efficient, relying only on vector inner products and norms, and thus introduces negligible overhead.

\noindent\textbf{Pixel Residuals Compression.} To enhance high-frequency details, and provide a solution for scalable bitrate control we introduce a pixel-residual computation mechanism.

On the encoder, as shown in Figure \ref{fig:system}, we employ a proxy model to convert encoded feature back to pixels in real time. Comparing the original frame $x$ with its reconstruction $\hat{x}$ yields a per-pixel residual $r = x - \hat{x}$ that precisely captures information lost during compression. However, directly transmitting raw residuals is impossible: for 1080p at 30\,fps, the residual stream is around 1.39\,Gbps, which is far beyond narrow-link budgets and at odds with our compression goals.

Therefore, we design a novel residual pipeline. First, we observe that residuals across neighboring frames are often similar, especially in static or slow-motion scenes. Exploiting this, we aggregate residuals over a temporal window of length $T$ by averaging:
\begin{equation}
r_{\text{compressed}}(h,w,c) \;=\; \frac{1}{T}\sum_{t=1}^{T} r(h,w,c,t)
\end{equation}
where \( h \) and \( w \) represent the height and width of the video and \( c \) represents the number of channels of the residuals. This temporal compression reduces both volume and noise residuals, since random noise diminishes under averaging. The compressed residual at each pixel is then distributed back to all frames in the window to ensure high visual quality.

Next,We leverage the sparse nature of residuals to achieve additional compression. Because VGC already preserves most of the salient content, the residual number are typically small. We apply thresholding with $\theta$, zeroing elements whose absolute value is below $\theta$, yielding a highly sparse residual matrix.

Finally, we use arithmetic entropy coding from traditional video codecs to losslessly compress the sparse residuals. Because of its high sparsity, arithmetic coding achieves excellent compression efficiency. Together, these steps compress residuals by nearly $2000\times$ within acceptable distortion, making transmission feasible even on narrow bandwidth networks.

\vspace{-0.5em}

\section{Resolution Scaling Accelerator}\label{downsample}
\vspace{-0.5em}
To address the computational inefficiency of original VFMs, this section presents a resolution scaling-based accelerator that improves system efficiency for our VGC.




\noindent\textbf{Adaptive Resolution Control.} Reducing input video resolution is the most direct way to lower both encoding compute and transmission bitrate. This offers multiple benefits: reduced bitrates make resolution control the main rate adjustment in \sys, enabling quick response to bandwidth changes, and lowers encoding and decoding complexity, improving efficiency on resource-limited devices. We develop an adaptive resolution control framework to achieve this.

However, the conventional downsample then super resolution (SR) pipeline faces two key challenges in semantic video coding. First, reconstructions from the VGC decoder often contain noise and artifacts, applying super resolution naively amplifies these degradations and harms perceptual quality. Second, because the codec is built on vision foundation models pretrained on high-quality data, feeding low-resolution inputs after downsampling degrades reconstruction performance and fails to deliver the desired visual quality.

To address these issues, we propose staged optimization where the codec is trained to produce outputs well-suited for super resolution, rather than forcing a lightweight SR model to adapt to codec artifacts. This design is driven by real-time constraints requiring lightweight SR networks with limited depth and parameters, which struggle to learn complex denoising mappings. In contrast, the codec's richer structure and capacity make it better positioned to learn how to emit outputs matching a target distribution.

The procedure employs a two-stage approach: Stage 1 independently trains a lightweight super resolution model on large-scale data augmented with random noise, compression artifacts, and quality degradations to establish robust priors and define the expected input distribution. Stage 2 freezes the SR weights and jointly fine-tunes the VGC on high-definition video to align its reconstructions with the SR model's training distribution, reframing the objective as distribution alignment rather than learning restoration mappings from scratch.

We perform joint fine-tuning end-to-end via the internal gradient interface in \S\ref{codec}. Because the super resolution model is already pretrained in simulated degradations, the codec’s gradient space is substantially reduced, making training more stable and efficient. This reverse-adaptation strategy improves the quality and compute trade-off. 

\vspace{-1em}
\section{Network-Adaptive Streaming Controller}\label{network}
\vspace{-0.5em}
In this section, we propose networking integration techniques for \sys, where NASC enables scalable bitrate control and robust streaming while achieving loss resilience.

\vspace{-1em}
\subsection{Scalable Bitrate Control}\label{control}
To accommodate dynamic network bandwidth, \sys achieves continuous and scalable bitrate adaptation. In this section, we first introduce how \sys controls the bitrate by coordinating three rate-control techniques: adaptive resolution control (\S\ref{downsample}), pixel-residual transmission and intelligent token dropping (\S\ref{drop and res}). Then, we describe how to adaptively select the target bitrate based on network bandwidth.

\noindent\textbf{Rate-control Framework.} \sys employs an anchor-based linear control scheme. We define two key anchors—the 3× downsampling anchor ($R_{3x}$) and the 2× downsampling anchor ($R_{2x}$)—which serve as thresholds for enabling different rate-control strategies. Given the measured available bandwidth $B_{\text{avail}}$, the controller selects a strategy bundle in real time, with relevant details available in the Appendix \ref{SBCal}.

In the extremely low bandwidth mode, the system encodes at 3× downsampling to encode tokens and then applies similarity-aware token selection from \S\ref{drop and res}, adapting the drop rate to the current bandwidth. In the low bandwidth mode, the system preserves the full 3× downsampled semantic representation and allocates the remaining bandwidth to pixel residuals from \S\ref{drop and res}, ensuring that salient regions receive higher-fidelity reconstruction. When bandwidth is sufficient, the controller switches to 2× downsampling for higher base quality and continues to devote surplus bandwidth to residual transmission to further enhance detail. To ensure smooth viewing, mode transitions use hysteresis to avoid oscillations due to bandwidth jitter. When the resolution changes, the system applies the temporal smoothing techniques from \S\ref{overlap} to preserve coherence across frames.

This hierarchical control strategy allows \sys to combine multiple mechanisms intelligently and to deliver better visual quality and bandwidth efficiency than conventional codecs across diverse network conditions (\S\ref{network evaluation}).

\noindent\textbf{Adaptive Bitrate Selection.} NASC adopts a receiver-driven control architecture and uses BBR (Bottleneck Bandwidth and Round-trip propagation time) \cite{cardwell2016bbr} for bandwidth estimation. The receiver reports the estimated available bandwidth every 100 ms, the sender then invokes the above bitrate controller to adjust encoding parameters on the fly.

\vspace{-1em}
\subsection{Robust Streaming Protocol}\label{stream}

Conventional video streaming systems adopt multiple packet loss resilience mechanisms, typically combining forward error correction (FEC), automatic repeat request (ARQ), and decoder-side error concealment \cite{wang1998error}. However, under adverse conditions these mechanisms often induce high latency or sharp swings in visual quality \cite{lambooij2009visual}.

Conventional codecs handle missing pixels through simple interpolation. In contrast, semantic codecs leverage high-level representations and can reconstruct missing content by exploiting surrounding semantic and spatiotemporal context. We therefore design a streaming protocol tailored to a semantic codec, tightly coupling transport with semantic reconstruction. The protocol comprises two core components: an intelligent packetization strategy, and a hybrid loss handling strategy. These components work in concert to deliver resilient video without injecting extra redundancy.

\begin{figure}[t]
    \centering
    \includegraphics[width=0.45\textwidth]{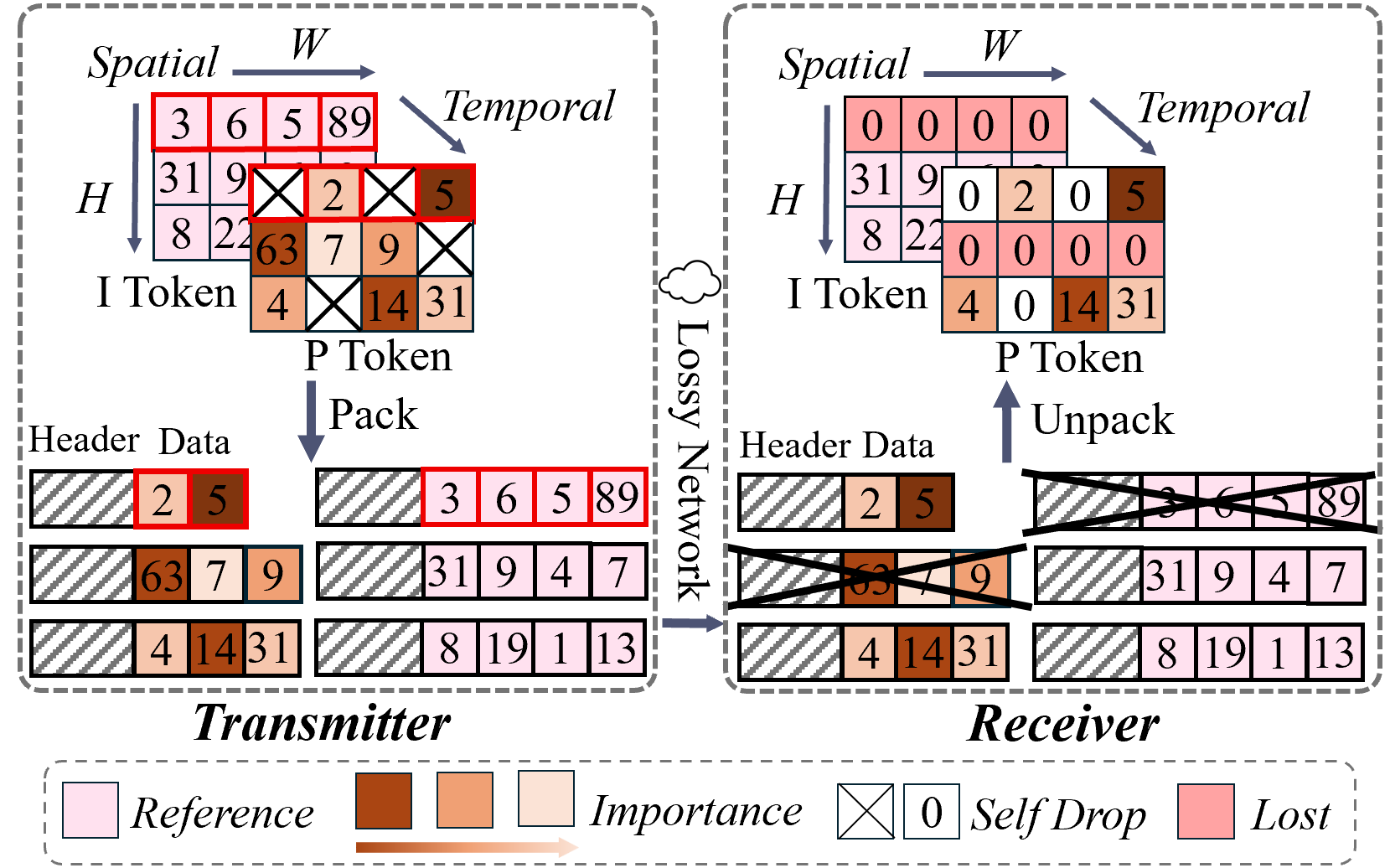}
    \caption{Token Packetization and Hybrid Loss Handling.}
    \label{fig:drop}
    \vspace{-1em}
\end{figure}

\noindent\textbf{Intelligent Packetization.} NASC introduces a token-oriented packetization scheme that reframes packet loss as a denoising problem the network has already learned to handle. As shown in Figure \ref{fig:drop}, for an $H \times W$ token matrix emitted by the encoder, we package row by row into $H$ packets of length $W$. Each packet carries a header and a payload of valid tokens. The header records the row index in the original matrix and a position mask for tokens in the packet, represented as a binary vector of length $W$ where one denotes a valid token and zero denotes a proactively dropped token.

A key innovation is the unified treatment of missing information. At the receiver, the decoder reassembles each received row using the header’s row index and mask, placing valid tokens at their locations and filling zeros at masked positions corresponding to proactively dropped tokens. If a packet is lost, the entire row is zero filled. From the decoder’s perspective, similarity-based token dropping and network loss appear identically as zero-valued noise.

Because \sys has been jointly trained to reconstruct from incomplete token matrices (\S\ref{implementation}), packet loss merely increases the amount of noise to process rather than introducing a new failure mode. The system leverages the codec’s intrinsic capability and does not require additional error-correction layers to remain robust.

\noindent\textbf{Hybrid Loss Design.} \sys differentiates loss policies for semantic tokens and residuals. For semantic tokens, the system decodes partial data directly and triggers retransmission only when the loss rate exceeds a preset threshold, typically 50\%. For residuals, the policy is looser. If a residual packet is lost, the corresponding frame simply skips residual enhancement rather than waiting for a resend. The basic insight is that semantic tokens carry the core content, whereas residuals primarily add detail. Under constrained or unstable networks, sustaining continuous delivery of core content is more important than perfectly reconstructing detail.

This content-aware streaming protocol yields graceful quality degradation instead of hard stalls when conditions worsen, substantially improving user experience. At the same time, because the system can operate on incomplete data, retransmissions are reduced, which lowers end-to-end latency and improves bandwidth utilization.

\vspace{-2mm}
\section{Implementation of \sys}\label{implementation}
\vspace{-0.5em}

\begin{figure}[t]
    \centering
    \includegraphics[width=0.42\textwidth]{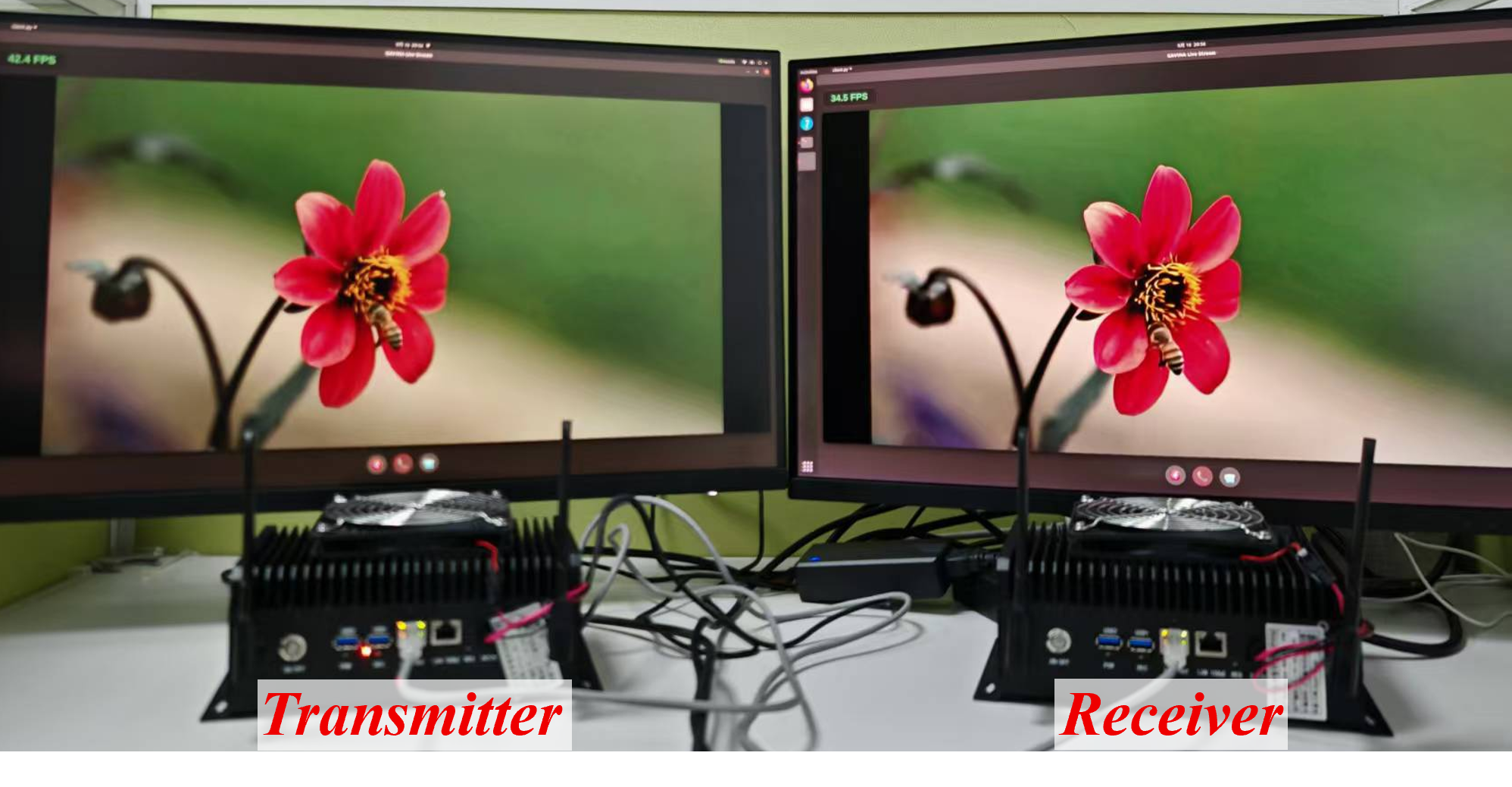}
    \caption{Showing Devices Used in System Prototype. Two NVIDIA Jetson serve as transmitter and receiver connected via network cable for end-to-end transmission.}
    \label{fig:implementation}
    \vspace{-2em}
\end{figure}

\sys is implemented with approximately 35K lines of Python and C++ code, including its codec model, scalable bitrate adaptation, streaming protocol control and a WebRTC-based prototype.

\noindent\textbf{Codec Training.} The \sys codec architecture builds on the recent Cosmos vision foundation model \cite{agarwal2025cosmos}. We fine-tune \sys from pretraining this VFM on the UltraVideo-Long dataset \cite{xue2025ultravideo} using a two-stage training strategy. 
Stage one trains the base model without packet loss to optimize inter-frame temporal smoothness and adaptive resolution support. Stage two introduces random token drop training (\S\ref{drop and res}) with drop rates sampled uniformly from $[0\%, 25\%]$. Unlike decoder-only error concealment, \sys employs joint training to align the encoder and decoder for handling token loss. The training objective combines multiple loss functions.\footnote{Please note that the $\ell_{1}$ norm between $\hat{x}$ and $x$ is strongly correlated with the $\mathrm{PSNR}$ of $\hat{x}$ (with respect to $x$). To avoid this bias, in our evaluation in \S\ref{evaluation} we report quality improvements using alternative metrics such as VMAF, etc.} We train VGC on 16 NVIDIA A100 GPUs consuming approximately 195 GPU hours total. Due to space limitations, we provide more details about \sys training in Appendix \ref{sec:appendix:training_topology}.

\begin{figure*}[ht!]
\vspace{-5mm}
    \centering
    \includegraphics[width=1\linewidth]{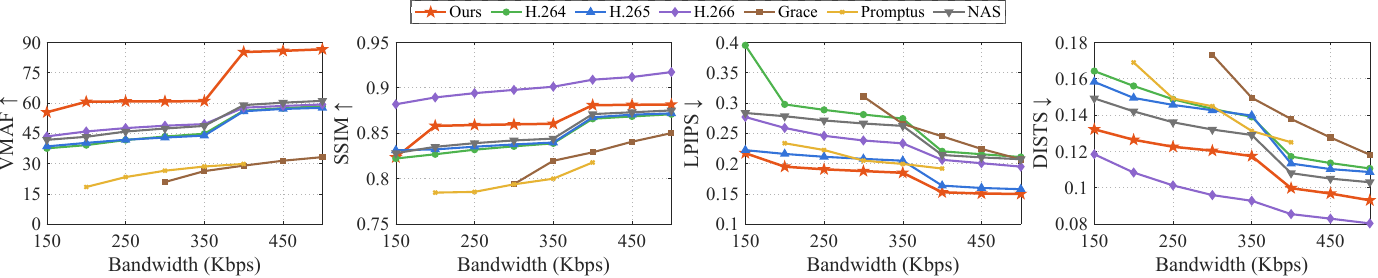}
    \vspace{-6mm}
    \caption{Rate-Distortion Performance of Our Generative Codec (UGC Dataset).} 
    \label{fig:rd_curve} 
\end{figure*}

\noindent\textbf{Codec Wrapper.} To enable end-to-end video streaming, we implement a wrapper that integrates the VGC into the transmission pipeline. We compile \sys with PyTorch JIT version 2.6 and use BFP16 for encoding and decoding computations. The wrapper maintains model parameter states at both sender and receiver, eliminating initialization overhead. For instance, when the encoder adjusts its downsampling factor, the decoder's super-resolution module can immediately switch to the corresponding model parameters without reloading network weights over PCIe.

\noindent\textbf{\sys Prototype.} We build a \sys prototype based on WebRTC \cite{webrtc2021}, running on two NVIDIA Jetson AGX Orin (32 GB) devices respectively. As shown in Figure \ref{fig:implementation}, one Jetson serves as the sender and the other as the receiver, each running the wrapped \sys codec model. During operation, the codec interacts with the scalable bitrate control (\S\ref{control}) module through the external data interface to enable dynamic rate adaptation. The two devices connect via network cable. To test our robust streaming protocol (\S\ref{stream}) in lossy network conditions, we also build a trace-based network emulator. This network emulator acts as a relay, receiving and forwarding packets between the sender and the receiver. In the emulator, packets are randomly dropped and mahimahi \cite{191577} is used to replay the real network traces from Puffer~\cite{yan2020learning}.  Both sender and receiver use QT to render the playback interface, which displays the real-time coding FPS in the top left corner.

\vspace{-3mm}
\section{Evaluation}
\label{evaluation}
\vspace{-1mm}
\subsection{Setup}

\textbf{Test Videos}: Our test dataset consists of 100 unique 10-second videos at 1080p resolution, including both 30 and 60 FPS (total 1000 seconds), carefully chosen from four public datasets: UVG \cite{mercat2020uvg}, UHD \cite{xue2025ultravideo}, UGC \cite{wang2019youtube}, and Inter4K \cite{stergiou2022adapool}. The comprehensive and diverse dataset allows us to test \sys's performance in real-life video streaming tasks.

\noindent\textbf{Baselines}: To comprehensively evaluate \sys's performance, we selected a set of representative video codecs as baselines. Among these, we first included H.264, H.265 and H.266\cite{bross2021overview}, which are the most widely adopted codecs (implemented via FFmpeg v7.1.3). Furthermore, we experimented with recent advancements such as Grace\cite{cheng2024grace}, a representative of neural codecs, and Diffusion Model based Promptus \cite{wu2024promptus}.

\noindent\textbf{Metrics}: Following the latest practices in video streaming performance evaluation \cite{de2013adaptive}, we select 3 aspects:
\begin{itemize}
    \item \textbf{Visual Fidelity} is measured using four main metrics: SSIM \cite{wang2004image}, VMAF \cite{rassool2017vmaf}, LPIPS \cite{zhang2018unreasonable}, and DISTS \cite{ding2020image}. To assess the temporal stability of neural streaming, we introduce inter-frame temporal consistency metrics. Our primary goal is to maintain a high-quality user viewing experience at low bitrates. These metrics collectively reflect human subjective perception quality for video playback performance.
    
    \item \textbf{Real-time performance} for \sys is evaluated by measuring the per-frame processing time for both the encoder and decoder, as well as the per-frame transmission delay under complex network conditions.

    \item \textbf{Video Loss-resilience} refers to the system's robustness and stability when facing packet loss conditions. We report both the visual quality under packet loss and the client-side rendered frame rate under poor network conditions.
\end{itemize}

\vspace{-1em}
\subsection{Evaluation of Visual Performance}

\textbf{Compression versus Visual Quality.}We conducted extensive experiments to validate that our system effectively balances the trade-off between aggressive compression and high fidelity reconstruction. Figure \ref{fig:rd_curve} demonstrates that our generative codec significantly outperforms traditional codecs and generative baselines across most metrics on the UGC dataset within the selected bandwidth range. Our method achieves a VMAF score of 85.17 compared to H.266's 57.61 and H.265's 55.85 at 400kbps, indicating superior perceptual quality under low bandwidth constraints. Although H.266 exhibits excellent performance in terms of SSIM and DISTS metrics, its computational complexity and the current limitations in hardware acceleration development constrain its practical efficiency. Notably, when transitioning from 350kbps to 400kbps through quality control mechanisms, the video downsampling ratio has been adjusted from 3x to 2x. Benefiting from efficient spatiotemporal compression and downsampling preprocessing, achieving over thousand-fold compression demonstrates promising prospects for applications.

\begin{figure}[t] 
    \centering
    \includegraphics[width=\linewidth]{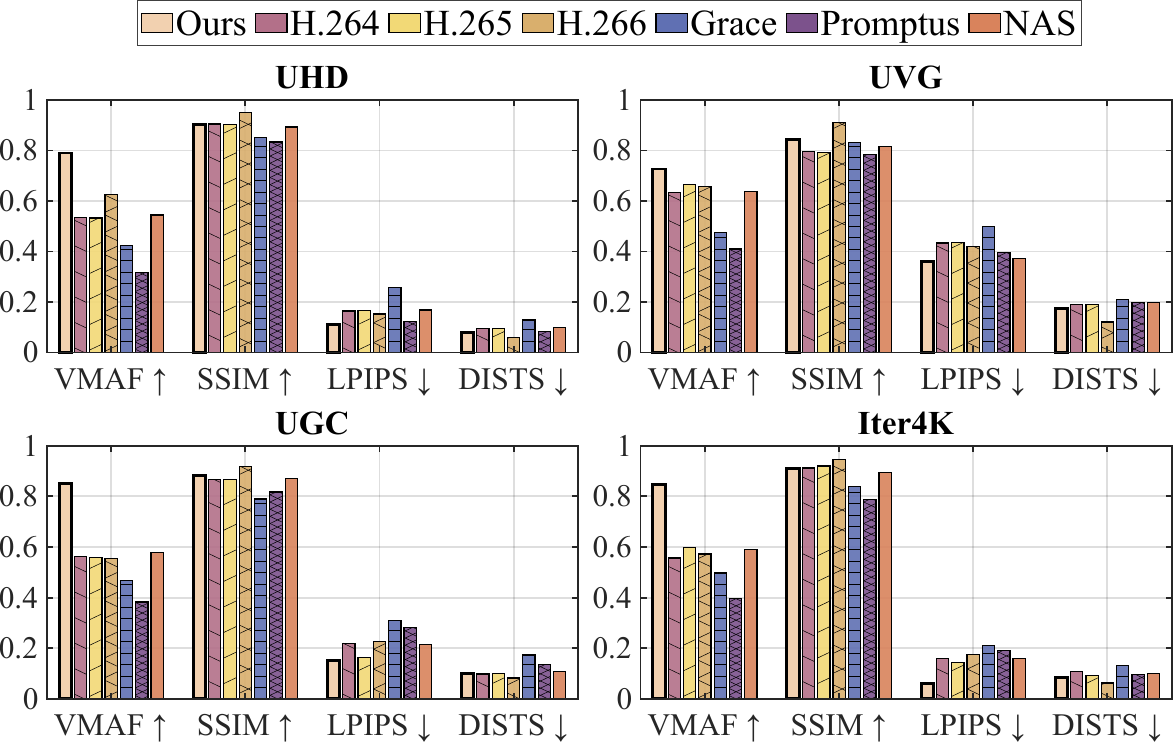}
    \vspace{-6mm}
    \caption{Visual Metrics in Four Datasets.} 
    \label{fig:metrics}
    \vspace{-2em}

\end{figure}

\noindent\textbf{\sys Generalizability.}   
We validated our model's performance across four diverse datasets. Figure \ref{fig:metrics} demonstrates our method's superior generalization capability, consistently achieving higher average VMAF scores compared to other methods while maintaining competitive performance with generative baselines across all perceptual metrics (SSIM, LPIPS, DISTS), thereby validating robust cross-domain applicability. We provide more details about the visual comparison of \sys and other baselines for video frames transmitted at 400kbps bitrate in Appendix \ref{appendix:visual}.

\begin{figure}[t]
  \centering
  \includegraphics[width=0.48\textwidth]{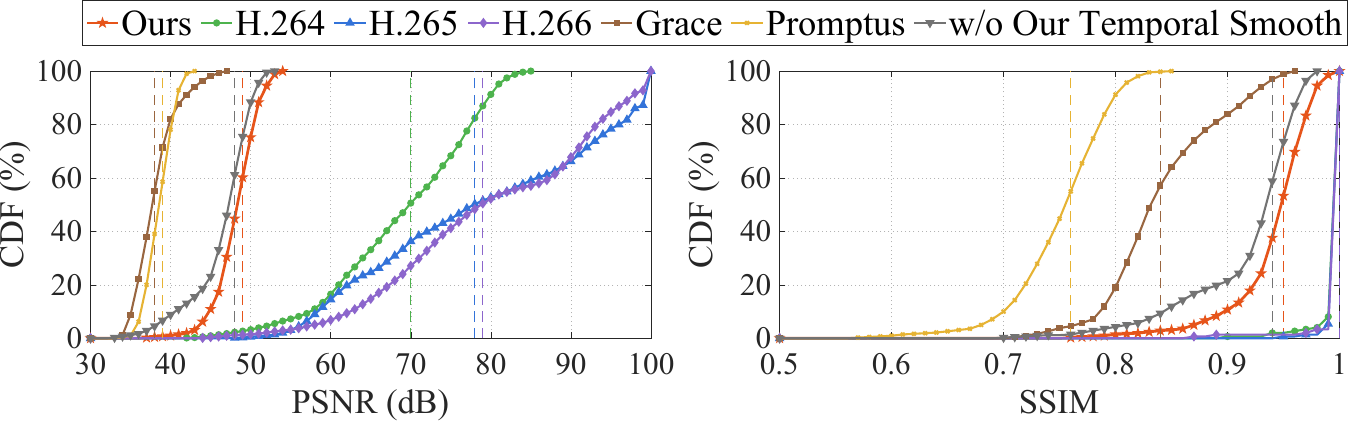}
  \vspace{-6mm}
  \caption{Comparison of Video Temporal Consistency.}
  \label{fig:clip}
  \vspace{-1mm}
\end{figure}

\noindent\textbf{Video Consistency.} As illustrated in Figure \ref{fig:clip}, we present the distribution curves of visual temporal consistency across two metrics. We compute the inter-frame residuals between consecutive frames of each videos and assess the quality by comparing them to the residuals of the original video. Analysis of pixel-level variations between consecutive frames reveals that all neural-based systems, including GRACE and Promptus, demonstrate considerably more temporal flickering than traditional codecs. To enhance the temporal coherence of generative models, we introduce a temporal smoothing technique that improves the visual perception experience. The proposed \sys system achieves superior pixel structure stability and consistency, substantially outperforming existing neural-based streaming methods.

\vspace{-2mm}
\subsection{Evaluation of Networking System}\label{network evaluation}

\begin{table}[t]
\centering
\caption{Computational Overhead for Different Devices}
\vspace{-1mm}
\renewcommand{\arraystretch}{0.8}  
\setlength{\tabcolsep}{1.5pt}      
\footnotesize                      
\begin{tabular}{ccccc}
\toprule
\textbf{Device} & Scale & GPU Memory (GB) & Encoder (FPS) & Decoder (FPS) \\
\midrule
\multirow{2}{*}{RTX3090} & 3x & 8.86 & 98.51 & 65.74 \\
 & 2x & 17.09 & 47.14 & 32.03 \\
\cmidrule(l){2-5}
\multirow{2}{*}{A100} & 3x & 7.96 & 101.23 & 83.33 \\
 & 2x & 16.24 & 52.54 & 40.19 \\
\cmidrule(l){2-5}
\multirow{2}{*}{Jetson} & 3x & 15.21 & 61.17 & 43.45 \\
 & 2x & 23.87 & 31.87 & 24.93 \\
\bottomrule
\end{tabular}
\label{tab:gpu_performance}
\vspace{-1.5em}
\end{table}

\noindent \textbf{Real-time Performance.} We conducted comprehensive experiments across different platforms, including Nvidia RTX 3090, A100, and Jetson Orin devices, with results shown in Table~\ref{tab:gpu_performance}. Our evaluation focuses on key streaming metrics including encoding/decoding frame rates, memory consumption, and scalability under different resolution scaling factors (2$\times$ and 3$\times$). The results demonstrate that our system achieves real-time performance across all platforms: the A100 delivers 101 FPS encoding and 83 FPS decoding at 3$\times$ scaling, while the RTX 3090 provides balanced performance suitable for consumer-grade streaming. Notably, even the resource-constrained Jetson Orin maintains practical frame rates (61/43 FPS at 3$\times$ scaling), validating our system's deployability on edge devices. These measurements demonstrate the system's capability for live video streaming and provide optimal deployment strategies for various streaming applications ranging from cloud servers to edge devices.

\begin{figure}[t] 
    \centering
    \includegraphics[width=\linewidth]{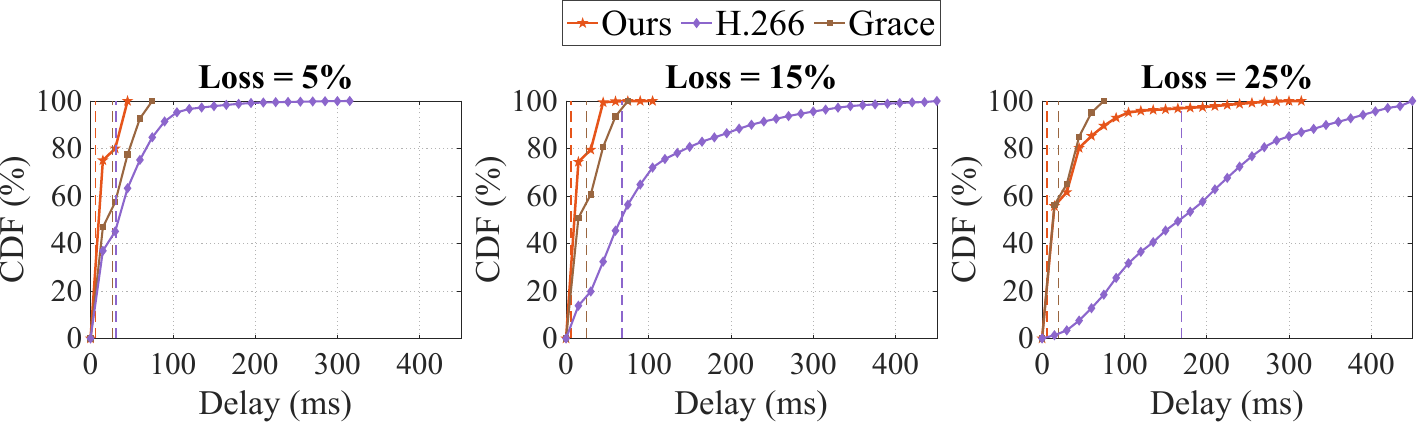}
    \vspace{-6mm}
    \caption{Frame Latency Distribution for Video Streaming at Different Packet Loss Rates. The dashed lines indicate the median values.}
    \label{fig:delay}
    \vspace{-4mm}
\end{figure}

\begin{figure}[t] 
    \centering
    \includegraphics[width=\linewidth]{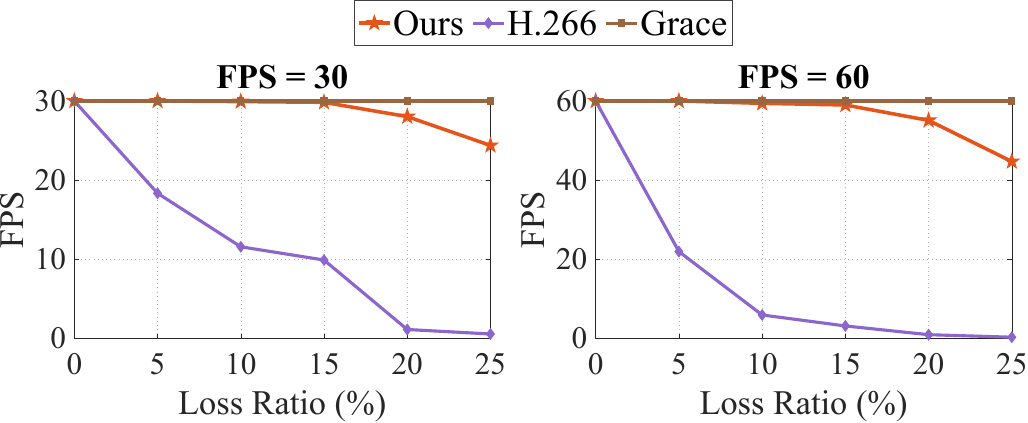}
    \vspace{-6mm}
    \caption{Comparison of Decoded and Rendered Frame Rates Between Videos Encoded at 30 fps and 60 fps. The decline in rendered frames suggests that packet loss has either prevented frame rendering entirely or caused substantial visual quality degradation.}
    \label{fig:renderedfps}
    \vspace{-3mm}
\end{figure}

\noindent \textbf{Loss Resilience.} To evaluate the loss resilience of the systems in complex network environment, we tested the frame delay and video stuttering (measured by average rendering frame rate \cite{rudow2023tambur}) under packet loss rates ranging from 5\% to 25\% at 400kbps. (\sys can tolerate packet loss up to 30\%, but we cap it at 25\% to ensure visual quality.) Figure \ref{fig:delay} demonstrates that our system maintains low latency even under severe packet loss conditions. At 25\% packet loss, \sys and Grace achieve sub-150ms delay for over 90\% of frames, while H.266 suffers substantial degradation with only 40\% of frames meeting this threshold. This validates the effectiveness of our loss-resilient streaming design. Figure \ref{fig:renderedfps} demonstrates the rendered frame rate performance under packet loss, measured by average FPS at two target frame rates. Our system and Grace maintain stable performance across all loss rates, sustaining near-target FPS even at 25\% loss. In contrast, H.266 suffers severe degradation, dropping to nearly 1 FPS at 25\% loss, rendering the video essentially unwatchable.

\begin{figure}[t] 
    \centering
    \includegraphics[width=\linewidth]{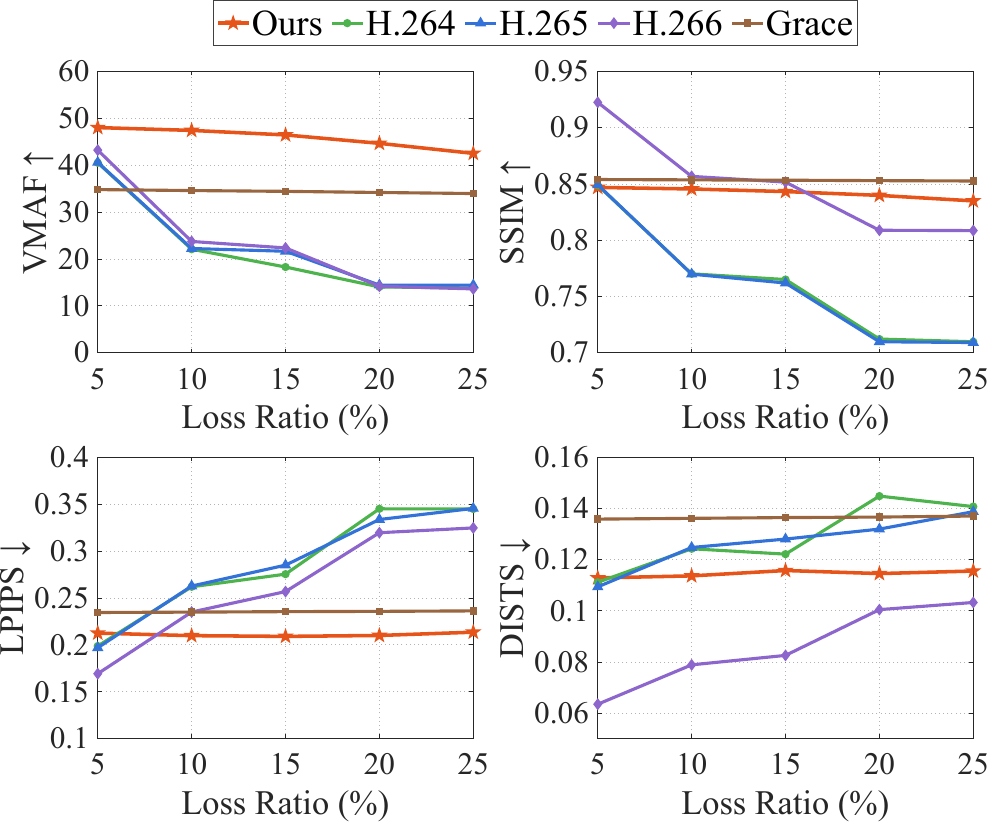}
    \vspace{-5mm}
    \caption{Visual Performance Benchmark under Different Packet Loss Rates.}
    \vspace{-3mm}
    \label{fig:loss_metrics}
\end{figure}

Beyond system transmission performance in complex network environments, this study evaluates visual quality under packet loss. We tested multiple visual metrics at 400kbps with packet loss rates from 5\% to 25\%, as shown in Figure \ref{fig:loss_metrics}. For key metrics measuring video semantic information and objective quality—VMAF, LPIPS, and DISTS—\sys not only achieved optimal performance but also exhibited minimal degradation, demonstrating strong stability. In contrast, traditional video codecs showed significant performance decline as packet loss increased. While the GRACE model degraded more gradually, its pixel-based nature resulted in inherently lower visual quality under poor networks.

By integrating semantic streaming with VFMs' generative capabilities, \sys achieves superior high-fidelity video restoration in low-bandwidth, lossy networks while maintaining low latency and smooth playback, ensuring excellent visual quality and user experience.

\begin{figure}[t]
  \centering
  \includegraphics[width=\linewidth]{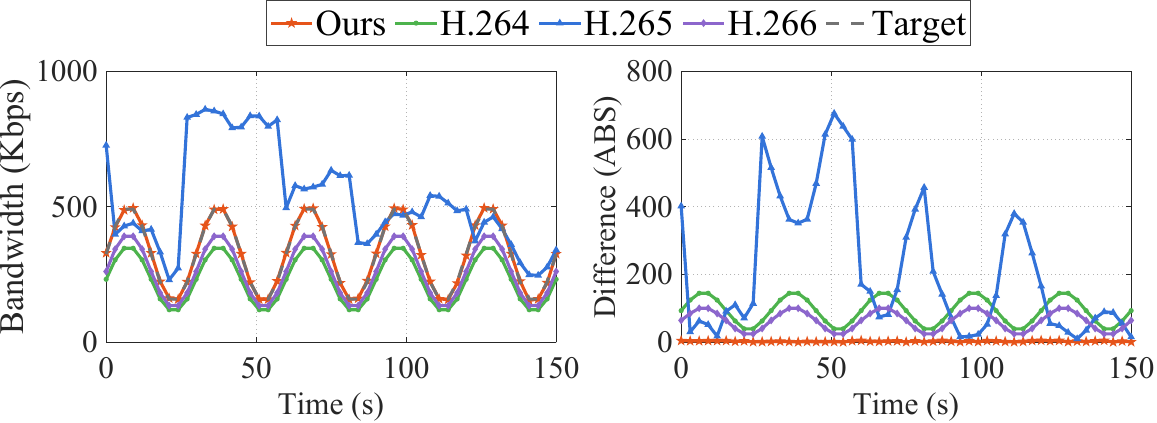}
  \vspace{-6mm}
  \caption{Network Bandwidth Trace Tracking Experiment.}
  \label{fig:trace}
\end{figure}

\noindent \textbf{Bitrate Control. }To evaluate system adaptation to dynamic network conditions, we used MahiMahi network emulator to simulate bitrate fluctuations between 200-500 kbps with 30-second periods. GRACE was excluded due to unavailable open-source bitrate control.
As shown in Figure \ref{fig:trace}, \sys demonstrated excellent bitrate tracking, with curve closely aligning with targets. This indicates effective scalable bitrate control that accurately adjusts to bandwidth changes, ensuring stable streaming under complex network conditions. In contrast, traditional encoders showed significant performance variations. H.264 and H.266 tracked the target relatively well but exhibited notable fluctuations and deviations. H.265 showed severe oscillations with large discrepancies from target bitrates, reaching up to 859.5 kbps during rapid changes -  causing network congestion and packet loss.

\vspace{-0.5em}

\begin{table}[t]
\centering
\caption{Ablation Study of Individual Module Contributions}
\vspace{-1mm}
\renewcommand{\arraystretch}{0.9}
\setlength{\tabcolsep}{3pt}
\footnotesize
\begin{tabular}{lccccc}
\toprule
\textbf{Method} & VMAF$\uparrow$ & SSIM$\uparrow$ & LPIPS$\downarrow$ & DISTS$\downarrow$ & Latency (ms) \\
\midrule
w/o \textit{RSA} & 59.72 & 0.84 & 0.22 & 0.14 & 644.61 / 874.76 \\
w/o Residual & 60.54 & 0.85 & 0.20 & 0.13 & \textbf{78.21} / \textbf{98.36} \\
w/o Self Drop & 20.31 & 0.73 & 0.41 & 0.23 & 89.82 / 136.87 \\
\sys & \textbf{60.76} & \textbf{0.86} & \textbf{0.18} & \textbf{0.11} & 91.36 / 136.91\\
\bottomrule
\end{tabular}
\vspace{1mm}

{\footnotesize \raggedright \textit{Note:} Latency values represent encoding/decoding latency (in milliseconds) for a chunk of 9 frames, respectively.\par}

\label{tab:ablation_study}
\vspace{-3mm}
\end{table}

\subsection{Ablation Study}
\vspace{-0.8mm}
Table \ref{tab:ablation_study} presents an ablation study evaluating each component's contribution to system performance, with latency in table referring to encoder/decoder delays. Removing Intelligent Self Drop (\S\ref{drop and res}) causes the most severe degradation, with VMAF dropping from 60.76 to 20.31 and LPIPS/DISTS deteriorating significantly, demonstrating its critical role in maintaining visual quality. The RSA (\S\ref{downsample}) module improves system efficiency by reducing latency, making computationally intensive AI models more suitable for real-time streaming. Although the Residual (\S\ref{drop and res}) incurs extra computational cost from the proxy model, it significantly enhances bandwidth utilization and enables finer-grained quality scheduling. Together, they validate the necessity of each component in balancing quality and latency trade-offs.

We tested \sys under 50\% packet loss using both Intelligent Self Drop and Random Drop strategies. As shown in Figure \ref{fig:sem_vs_rd} in the Appendix \ref{appendix:abs}, Intelligent Drop outperformed Random Drop in both visual quality and quantitative metrics. The semantic importance-based self-drop mechanism effectively minimized visual degradation in critical regions.

To evaluate the effectiveness of temporal smoothing, we conduct temporal video segment analysis and visualization. As shown in Figure~\ref{fig:overlap} in Appendix~\ref{appendix:abs} and the quantitative results in Figure~\ref{fig:clip}, the system without this mechanism exhibits significantly greater temporal flickering effects, severely degrading the viewing experience.

\vspace{-3mm}
\section{Conclusion and Discussion}
\vspace{-1mm}
We present \sys, the first VFM-powered video streaming system delivering high-fidelity, low-latency transmission under challenging network conditions. Our system achieves a 62.5\% bitrate reduction compared to H.265 while maintaining comparable visual quality, with real-time streaming and 94.2\% bandwidth utilization in live deployments.

Our VFM-driven approach has trade-offs: fine-grained details like small signage may be lost, and text-heavy content remains challenging due to limited textual training in current VFMs—a common limitation across generative streaming methods. Despite these constraints, \sys represents a significant advance in video streaming, demonstrating that VFM-based video delivery is ready for real-world applications.

\vspace{-4mm}
\section*{Acknowledgments}
\vspace{-2mm}
We sincerely thank our shepherd Junchen Jiang, anonymous reviewers for their valuable feedback. The work was supported in part by the National Natural Science Foundation of China (Grant No. 62293482), the Basic Research Project No. HZQB-KCZYZ-2021067 of Hetao Shenzhen-HK S\&T Cooperation Zone, the National Key Research and Development Program of China (Grant No. 2024YFB2907000), the National Natural Science Foundation of China (Grant No. 62471423), the Shenzhen Science and Technology Program (Grant No. JCYJ20241202124021028 and Grant No. JCYJ20230807114204010), the Guangdong Talents Program (Grant No. 2024TQ08X346), the Shenzhen Outstanding Talents Training Fund 202002, the Young Elite Scientists Sponsorship Program of CAST (Grant No. 2022QNRC001), the Guangdong Provincial Key Laboratory of Future Networks of Intelligence (Grant No. 2022B1212010001) and the Shenzhen Key Laboratory of Big Data and Artificial Intelligence (Grant No. SYSPG20241211173853027). The work of Xinggong Zhang and Jiangkai Wu was supported by the National Natural Science Foundation of China (Grant No. 62431017).





\bibliographystyle{plain}
\bibliography{reference}
\appendix
\section{Appendix}
\label{sec:appendix}

\subsection{Scalable Bitrate Control Algorithm}\label{SBCal}
Algorithm \ref{alg:msbc} presents \sys's bitrate control procedure. The \textsc{ComputeResidual} and \textsc{TokenSelection} routines are from the scalable video coding techniques described in \S\ref{drop and res}.

\begin{algorithm}[h]
\caption{Scalable Bitrate Control}
\label{alg:msbc}
\begin{algorithmic}[1]
\REQUIRE Available bandwidth $B_{\text{avail}}$, video frame $F$
\ENSURE Encoded bitstream
\IF{$B_{\text{avail}} < R_{3x}$}
    \STATE $y \leftarrow \textsc{Encode\_3x}(F)$
    \STATE $y_{\text{filtered}} \leftarrow \textsc{TokenSelection}(y, B_{\text{avail}})$
    \STATE \textbf{return} $y_{\text{filtered}}$
\ELSIF{$R_{3x} \le B_{\text{avail}} < R_{2x}$}
    \STATE $y \leftarrow \textsc{Encode\_3x}(F)$
    \STATE $\textit{residual} \leftarrow \textsc{ComputeResidual}(F, y, B_{\text{avail}} - R_{3x})$
    \STATE \textbf{return} $\{y, \textit{residual}\}$
\ELSE
    \STATE $y \leftarrow \textsc{Encode\_2x}(F)$
    \STATE $\textit{residual} \leftarrow \textsc{ComputeResidual}(F, y, B_{\text{avail}} - R_{2x})$
    \STATE \textbf{return} $\{y, \textit{residual}\}$
\ENDIF
\end{algorithmic}
\end{algorithm}

\begin{figure*}[t]
  \centering
  \includegraphics[width=1\textwidth]{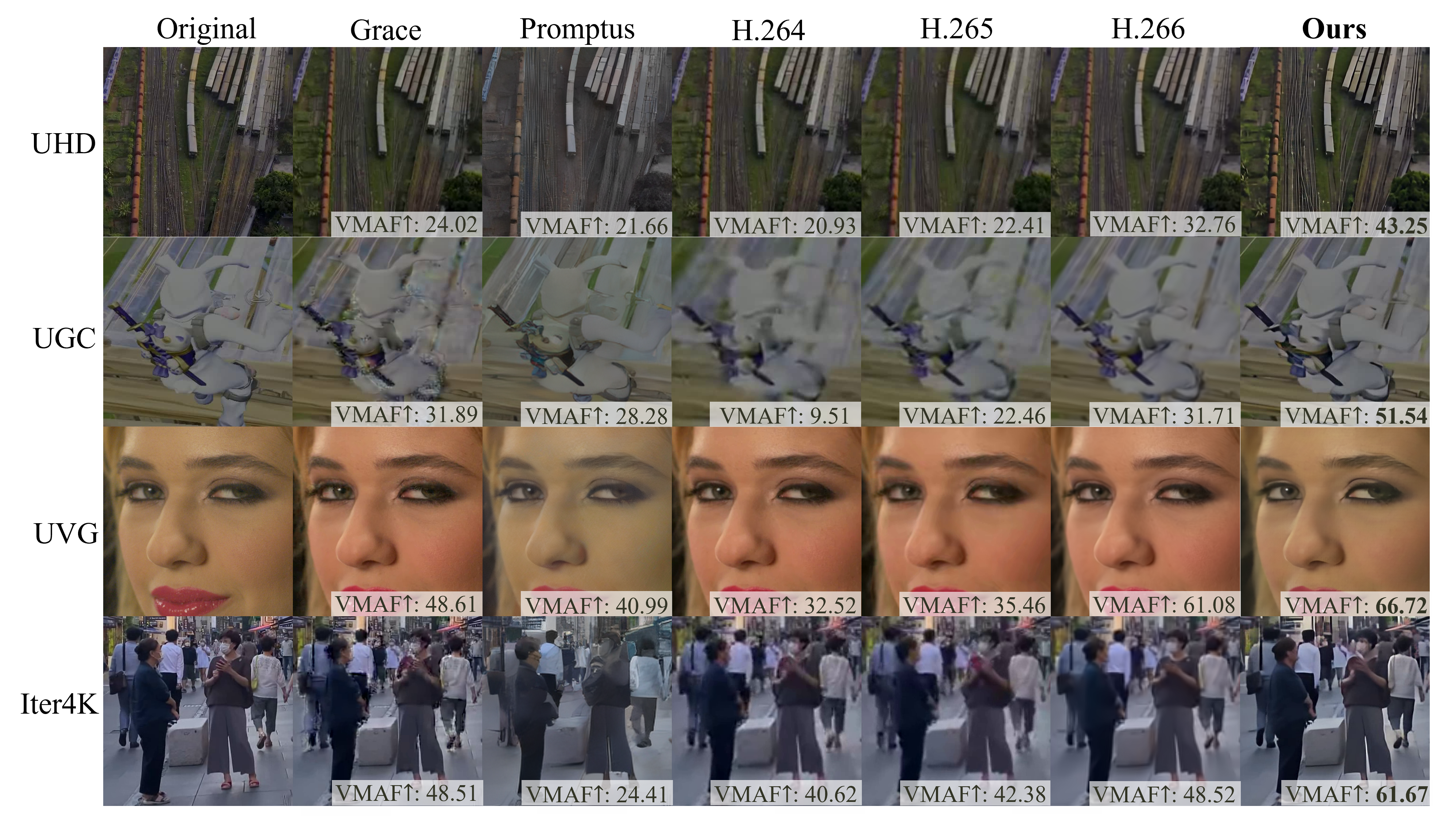}
  \caption{Visual quality comparison on four datasets.}
  \label{fig:vis_compare}
\end{figure*}

\subsection{\sys Training Flow}
\label{sec:appendix:training_topology}

\noindent\textbf{Base Codec Training Framework.} \sys fine-tunes a pretrained visual foundation model (VFM) backbone via the internal gradient interface described in \S\ref{codec}. The first training stage focuses on two core capabilities: inter-chunk temporal smoothing (\S\ref{overlap}) and adaptive resolution control (\S\ref{downsample}). The optimization objective is defined as:

\begin{equation}
L = \alpha L_{\text{pixel}}(\hat{x}, x) + (1-\alpha) L_{\text{flow}}(\hat{x}, x)
\label{equal:L}
\end{equation}

where $L_{\text{pixel}}$ denotes the $\ell_1$ pixel reconstruction loss, $L_{\text{flow}}$ denotes the optical flow consistency loss, and the weighting parameter $\alpha$ is set to 0.8. The flow loss encourages temporal coherence across Group of Pictures (GoP) boundaries, while the pixel loss ensures spatial detail preservation.

To accommodate adaptive video encoding, the training data is augmented by resizing the original 1080P videos to 540P and 720P resolutions. During training, video frames of varying resolutions are randomly sampled to enhance the model's adaptability to multi-resolution inputs.

Furthermore, an adversarial GAN loss \cite{zhang2024degradation} term $L_{\text{adv}}$ is incorporated to enrich video detail fidelity. A hyperparameter $\gamma=0.1$ balances the contribution of the GAN loss. This adversarial objective promotes the generation of visually plausible details, enabling the model to capture finer and more natural textures and dynamic patterns.

Accordingly, the overall training loss is formulated as:

\begin{equation}
\label{equal:gan loss}
L_{\text{total}} = L + \gamma \cdot L_{\text{adv}}
\end{equation}

This training scheme allows \sys to maintain spatiotemporal consistency while flexibly supporting multi-resolution video inference, significantly enhancing the richness and realism of generated video details.

\noindent\textbf{Video Supper Resolution Model Training.} We first use the Codec model trained in the previous stage to encode 540P videos at the lowest bitrate. This encoding and decoding process introduces some loss and compression artifacts. The resulting degraded videos form the training dataset for the super-resolution model. The super resolution (SR) model is built using a residual convolutional network \cite{he2016identity}. It is initially trained on this dataset to learn how to restore the original videos from blurred, jittered, and compressed inputs. The loss during this stage follows the same formulation as in Equation~\eqref{equal:L}, which combines pixel reconstruction and optical flow consistency losses to preserve spatial details and temporal coherence. After pretraining, the weights of the SR model are fused with the decoder. We freeze the SR model weights and only fine-tune the codec weights. During this joint training, the original 1080P videos are linearly downsampled by random scales of 2x or 3x. These low-resolution frames are fed into the codec model for inference, producing super-resolved videos. To further improve SR performance, the output videos are supervised with the corresponding original high-resolution videos using the GAN loss defined in Equation~\eqref{equal:gan loss}. The training process leverages the internal gradient interface introduced in \S\ref{codec} for gradient backpropagation and iterative optimization. By freezing the SR model and only updating the codec weights, the encoder learns to generate pixel features that match the expected distribution of the SR model. This helps enhance the SR restoration quality and final video fidelity.

\noindent\textbf{Robustness Training via Random Drop.} After establishing basic codec capability, stage two introduces token-drop training to enhance loss resilience.

For a P-frame semantic tensor $y \in \mathbb{R}^{H \times W \times T}$, given similarity scores $S_{i,j}$ and a target drop rate, we selectively discard high-similarity tokens. Dropped positions are zero filled to form a partial representation $\tilde{y}$. The decoder $g_{\theta}(\cdot)$ consumes $\tilde{y}$ and learns to reconstruct the original video, using the same objective as stage one:
\begin{equation}
L_{\text{robust}} \;=\; \alpha \, L_{\text{pixel}}\!\big(g_{\theta}(\tilde{y}), x\big) \;+\; (1-\alpha) \, L_{\text{flow}}\!\big(g_{\theta}(\tilde{y}), x\big)
\label{equal:L_robust}
\end{equation}

Crucially, via the internal gradient interface, gradients backpropagate into the encoder so it learns to emit semantic representations that are robust to token loss. During training, we first perform inference at the lowest bitrate. In this process, we apply a similarity-based token dropping method \S\ref{drop and res} to simulate autonomous packet loss. To adapt to varying network packet loss conditions, we uniformly sample the drop rate from the range $[0, 25\%]$. This equips the model with the ability to handle different levels of missing information.

Why is \sys more resilient to loss? Unlike decoder-only error concealment, joint training aligns the encoder and decoder to cope with token drops. First, the decoder learns to exploit reference information in the I-frame semantic matrix to infer and complete missing tokens in P frames. Second, the encoder learns to organize the semantic space so that redundant content shared by I and P frames lies closer, which improves the accuracy of similarity estimation.

This co-optimization has a direct benefit under bandwidth constraints. The system can safely discard high-similarity tokens that are redundant and prioritize lower-similarity tokens that carry novel information. Because the encoder imposes a more structured semantic organization, the decoder can more accurately recover dropped P-frame content from retained I-frame information. This adaptive allocation, coupled with cooperative training, allows \sys to maintain acceptable visual quality even with token loss rates up to 30\%, achieving robustness at the semantic and pixel layer. 

Throughout the entire training process, we use learning rates that decay gradually from $1 \times 10^{-5}$ to $2 \times 10^{-8}$. This learning rate schedule helps stabilize training by allowing larger parameter updates in the early stages for faster convergence, while progressively reducing the update magnitude towards the end to fine-tune the model and avoid overshooting. As a result, the model can achieve better generalization and improved performance.



\subsection{Details of Visual Comparison}
\label{appendix:visual}

\noindent\textbf{Visual quality comparison.} Figure \ref{fig:vis_compare} compares the visual quality of \sys against four baseline methods across different datasets. We evaluate each method on four representative datasets. All methods are compared at the same bitrate (400Kbps) using VMAF as the perceptual quality metric. \sys achieves the highest VMAF scores across all test scenarios. Note that while the \sys codec can encode and decode videos of arbitrary sizes and dimensions, we crop 1080p videos to square format for presentation clarity. The visual comparisons show that \sys maintains significant advantages even on challenging user generated content like UGC, demonstrating its robustness to diverse video types. Traditional codecs H.264, H.265 and H.266 perform worst across all tests. The neural codec Grace outperforms traditional methods but still lags considerably behind ours. For YUV videos in the UVG dataset, all methods except ours introduce severe color shifts that degrade viewing experience.

Traditional encoders exhibit uniform quality degradation with perceptible compression artifacts and blocking effects. GRACE shows significant mosaic distortions between moving objects and backgrounds, while Promptus, though preserving fine details, suffers from semantic inconsistencies due to diffusion model limitations.

\subsection{Ablation Study}
\label{appendix:abs}
\begin{figure}[t]
\vspace{-5mm}
  \centering
  \includegraphics[width=0.5\textwidth]{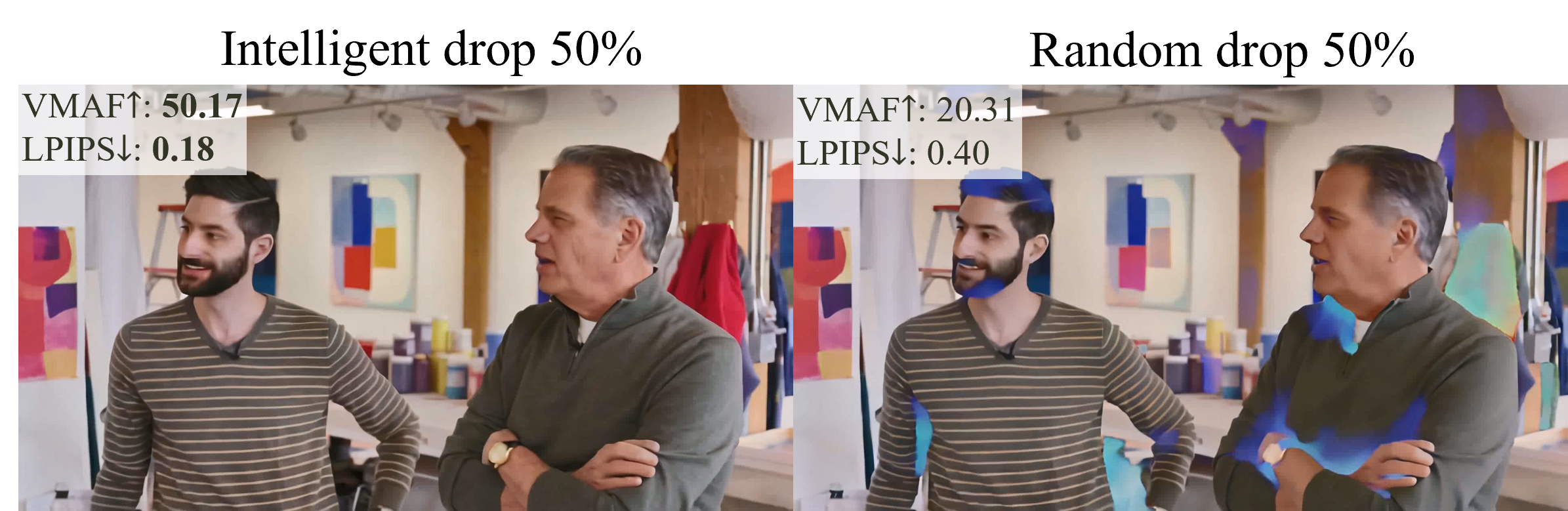}
  \vspace{-1mm}
  \caption{Our Token Dropping Solution vs Random Drop.}
  \label{fig:sem_vs_rd}
  \vspace{-3mm}
\end{figure}

To evaluate the efficacy of our intelligent token dropping mechanism, we compare it against naive random dropping at 50\% reduction rate in Figure~\ref{fig:sem_vs_rd}. The results reveal substantial performance differences: our method achieves VMAF of 50.17 and LPIPS of 0.18, whereas random dropping yields VMAF of only 20.31 and LPIPS of 0.40. The visual comparison clearly demonstrates that random dropping introduces severe corruption artifacts, particularly affecting texture-rich regions such as clothing and background details. Conversely, our intelligent approach maintains perceptual coherence by prioritizing semantically salient tokens, achieving approximately 2.5× higher VMAF scores and 55\% lower perceptual distortion. These results underscore the importance of content-aware token selection for quality-constrained video transmission.
\begin{figure}[t]
  \centering
  \includegraphics[width=0.5\textwidth]{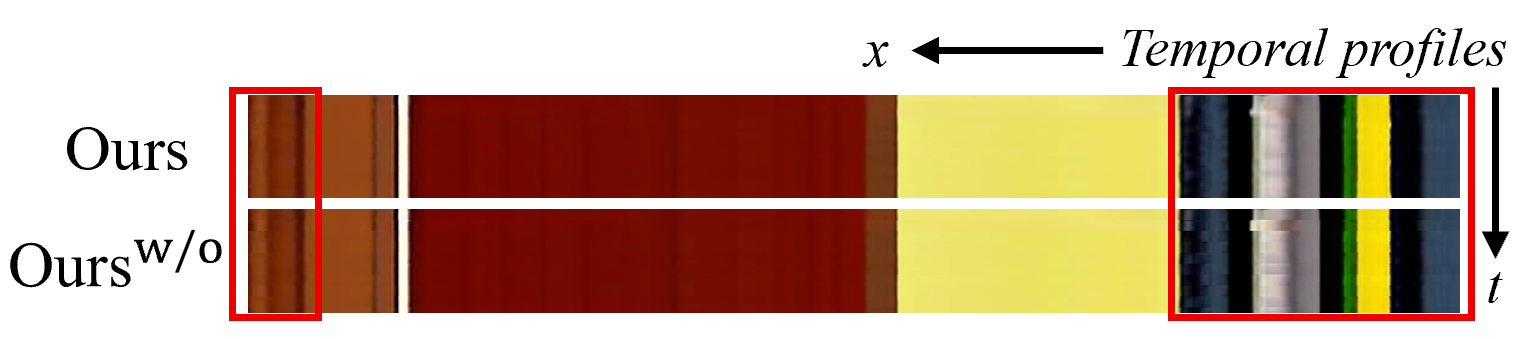}
  \vspace{-1mm}
  \caption{Ablation Study of Temporal Smoothing}
  \label{fig:overlap}
\end{figure}

Figure~\ref{fig:overlap} presents a visual comparison of temporal consistency with and without our smoothing mechanism. The temporal profile visualization reveals that the system without temporal smoothing (Ours w/o) exhibits significantly more color variation and discontinuities across consecutive frames, as evidenced by the pronounced fluctuations in the profile bands. In contrast, our complete system demonstrates substantially smoother transitions between frames, with more consistent color patterns throughout the sequence. This reduction in frame-to-frame variation directly translates to reduced flickering artifacts and improved viewing experience.

\end{document}